\documentclass[twocolumn]{aastex631}

\usepackage{comment}
\usepackage{rotating}
\usepackage{threeparttablex, tablefootnote}
\usepackage{xcolor}
\defcitealias{2023arXiv230104191L}{Lustig-Yaeger \& Fu et al. 2023}
\defcitealias{2023ApJ...948L..11M}{Moran \& Stevenson et al. 2023}
\defcitealias{May2023}{May \& MacDonald et al. 2023}

\begin{document}

\title{JWST COMPASS: NIRSpec/G395H Transmission Observations of TOI-776~c, a 2 R$_{\oplus}$ M Dwarf Planet} 

%tier 1: paper champion, theory lead JOINT FIRST AUTHORS

\author[0009-0008-2801-5040]{Johanna Teske$^{*}$} 
\affiliation{Earth and Planets Laboratory, Carnegie Institution for Science, 5241 Broad Branch Road, NW, Washington, DC 20015, USA}
\affiliation{The Observatories of the Carnegie Institution for Science, 813 Santa Barbara St., Pasadena, CA 91101, USA}
\thanks{These authors contributed equally to this work.}

\author[0000-0003-1240-6844]{Natasha E. Batalha$^{*}$}
\affiliation{NASA Ames Research Center, Moffett Field, CA 94035, USA}
\thanks{These authors contributed equally to this work.}

%%%I think this is a reverse alphabetical-order paper? 

%%%%%tier 2: data reductions that required iterations over the course of paper development%%%%%
\author[0000-0003-0354-0187]{Nicole L. Wallack}
\affiliation{Earth and Planets Laboratory, Carnegie Institution for Science, 5241 Broad Branch Road, NW, Washington, DC 20015, USA}

\author[0000-0002-4207-6615]{James Kirk}
\affiliation{Department of Physics, Imperial College London, Prince Consort Road, London, SW7 2AZ, UK}

%%%%%%tier 3: contributed a data reduction or modeling product(s), wrote text, synthesized results%%%%%%

\author[0000-0002-0413-3308]{Nicholas F. Wogan}
\affiliation{Space Science Division, NASA Ames Research Center, Moffett Field, CA 94035, USA}

\author[0000-0001-5253-1987]{Tyler A. Gordon }
\affiliation{Department of Astronomy and Astrophysics, University of California, Santa Cruz, CA 95064, USA}

\author[0000-0003-4157-832X]{Munazza K. Alam}
\affiliation{Space Telescope Science Institute, 3700 San Martin Drive, Baltimore, MD 21218, USA}

\author[0000-0002-8949-5956]{Artyom Aguichine}
\affiliation{Department of Astronomy and Astrophysics, University of California, Santa Cruz, CA 95064, USA}

%%%%%%%tier 4: contributed to paper comments or have otherwise contributed significantly to program%%%%%%

\author[0000-0003-2862-6278]{Angie Wolfgang}
\affiliation{Eureka Scientific Inc., 2452 Delmer Street Suite 100, Oakland, CA 94602-3017, USA}

\author[0000-0003-4328-3867]{Hannah R. Wakeford} 
\affiliation{School of Physics, University of Bristol, HH Wills Physics Laboratory, Tyndall Avenue, Bristol BS8 1TL, UK}

\author[0000-0003-3623-7280]{Nicholas Scarsdale}
\affiliation{Department of Astronomy and Astrophysics, University of California, Santa Cruz, CA 95064, USA}

\author[0000-0002-4489-3168]{Jea Adams Redai} 
\affiliation{Center for Astrophysics ${\rm \mid}$ Harvard {\rm \&} Smithsonian, 60 Garden St, Cambridge, MA 02138, USA}

\author[0000-0002-6721-3284]{Sarah E. Moran}
\affiliation{NASA Goddard Space Flight Center, 8800 Greenbelt Rd, Greenbelt, MD 20771, USA}
\affiliation{NHFP Sagan Fellow}

\author[0000-0003-3204-8183]{Mercedes L\'opez-Morales} 
\affiliation{Space Telescope Science Institute, 3700 San Martin Drive, Baltimore, MD 21218, USA}

\author[0000-0002-7500-7173]{Annabella Meech}
\affiliation{Center for Astrophysics ${\rm \mid}$ Harvard {\rm \&} Smithsonian, 60 Garden St, Cambridge, MA 02138, USA}

\author[0000-0002-8518-9601]{Peter Gao}
\affiliation{Earth and Planets Laboratory, Carnegie Institution for Science, 5241 Broad Branch Road, NW, Washington, DC 20015, USA}

\author[0000-0002-7030-9519]{Natalie M. Batalha}
\affiliation{Department of Astronomy and Astrophysics, University of California, Santa Cruz, CA 95064, USA}

\author[0000-0001-8703-7751]{Lili Alderson} 
\affiliation{Department of Astronomy, Cornell University, 122 Sciences Drive, Ithaca, NY 14853, USA}

 \author[0009-0003-2576-9422]{Anna Gagnebin} 
 \affiliation{Department of Astronomy and Astrophysics, University of California, Santa Cruz, CA 95064, USA}

%\author{et al.}
%\affiliation{Affiliations}

\begin{abstract}

The atmospheres of planets between the size of Earth and Neptune at short orbital periods have been under intense scrutiny. Of the $\sim$dozen planets in this regime with atmospheres studied so far, a few appear to have prominent molecular features while others appear relatively void of detectable atmospheres. Further work is therefore needed to understand the atmospheres of these planets, starting with observing a larger sample. To this end, %in this paper 
we present the 3--5\,$\mu$m transmission spectrum of TOI-776~c, a warm ($T_{\rm{eq}} \sim$420~K), $\sim$2~R$_{\oplus}$, $\sim$7~M$_{\oplus}$  planet orbiting an M1V star, measured with JWST NIRSpec/G395H. By combining two visits, we measure a median transit precision of $\sim$18 ppm and $\sim$32 ppm in the NRS1 and NRS2 detectors, respectively. We compare the transmission spectrum to both non-physical and physical models, and find no strong evidence for molecular features. For cloud-top pressures larger than 10$^{-3}$ bar, we rule out atmospheric metallicities less than 180-240$\times$ solar (depending on the reduction and modeling technique), which corresponds to a mean molecular weight of $\sim$6-8 g/mol. However, we find simple atmosphere mixture models (H$_2$O$+$H$_2$/He or CO$_2+$H$_2$/He) give more pessimistic constraints, and caution that mean molecular weight inferences are model dependent. We compare TOI-776~c to the similar planet TOI-270~d, and discuss possible options for further constraining TOI-776~c's atmospheric composition. Overall, we suggest these TOI-776~c observations may represent a combination of planetary and stellar parameters that fall just below the threshold of detectable features in small planet spectra; finding this boundary is one of the main goals of the COMPASS program.

\end{abstract}

% Reference Management 
% Public ADS LIbrary Link 
% https://ui.adsabs.harvard.edu/public-libraries/aHGglSvkQl-tJ7378bbvzg

%Classical Novae (251) --- Ultraviolet astronomy(1736) --- History of astronomy(1868) --- Interdisciplinary astronomy(804)}

\section{Introduction}
Exoplanets with sizes between 1-3~R$_{\oplus}$ and orbital periods $<$100 days are the most prevalent type of planet detected thus far \citep[e.g.,][]{Howard2012,Dressing2013, Fressin2013,Fulton2017}. However, these planets are absent from our own Solar System, and their origin is still mysterious. Thus, a major area of focus has been to determine their compositions and to constrain their formation pathway(s). 
In particular, several authors have suggested that some 1--3~$R_{\oplus}$ planets are 
water-rich based on a limited sample with bulk density measurements \citep{Zeng2019PNAS..116.9723Z,Mousis2020ApJ...896L..22M,Luque2022Sci...377.1211L}. \citet{Luque2022Sci...377.1211L} showed that, within the sample of precisely measured small planets discovered around M dwarf stars, 
there is a subset with densities consistent with a 1:1 rock:water composition. These so-called ``water worlds'' would therefore be a different population from rocky planets (which have  a larger range of densities) and ``puffy'' planets (which have densities consistent with rocky cores surrounded by massive H$_2$/He envelopes). However, the densities of the so-called ``water worlds'' can also be explained by planets with rocky interiors and H$_2$/He envelopes that undergo thermal evolution and mass loss \citep[e.g.,][]{Rogers2023} with no need for a ``water world'' population. 
With the true nature of planets between 1-3~R$_{\oplus}$ is not well understood, characterization of these planets' atmospheres offer an avenue to help break these compositional degeneracies.

With this overall goal in mind, we initiated the JWST COMPASS (Compositions of Mini-Planet Atmospheres for Statistical Study) Program in Cycle 1 (PID \# 2512, \citealt{2021jwst.prop.2512B}) 
to conduct a chemical inventory of a dozen small planet atmospheres. More details about 
COMPASS can be found in \cite{Batalha2023}, and 
the results published 
thus far
\citep{Alderson2024, Wallack2024, Scarsdale2024,Alam2024, Alderson2025}.

In this work, we present and interpret the observations of  
the COMPASS target TOI-776~c (TOI-776.01).  
TOI-776~c ($R_p=2.02\pm0.14$~R$_{\oplus}$; $M_p=5.3\pm1.8$~M$_{\oplus}$, $P = 15.67$~d) is one of two known planets orbiting the nearby, 
bright 
M1 dwarf star LP 961-53. The other known planet, TOI-776~b ($R_p=1.85\pm0.13$~R$_{\oplus}$; $M_p=4.0\pm0.9$~M$_{\oplus}$) orbits the host star with a period of 8.25~d \citep{Luque2021}.
Recently, \cite{Fridlund2024} published additional photometric (TESS \citep{Ricker2015} and CHEOPS \citep{Benz2021}) and spectroscopic (HARPS \citep{Mayor2003}) observations of the system, with updated stellar and planet parameters that include more precise planet radii. The two planets 
are similar in density ($\rho_b = 4.8^{+1.8}_{-1.6}$ vs. $\rho_c = 4.4^{+1.8}_{-1.6}$~g~cm$^{-3}$), with planet b 
having a slightly more precise mass.  
See Table \ref{table:system} for a full list of properties.

\begin{figure*}
    \centering
    \includegraphics[width=\textwidth]{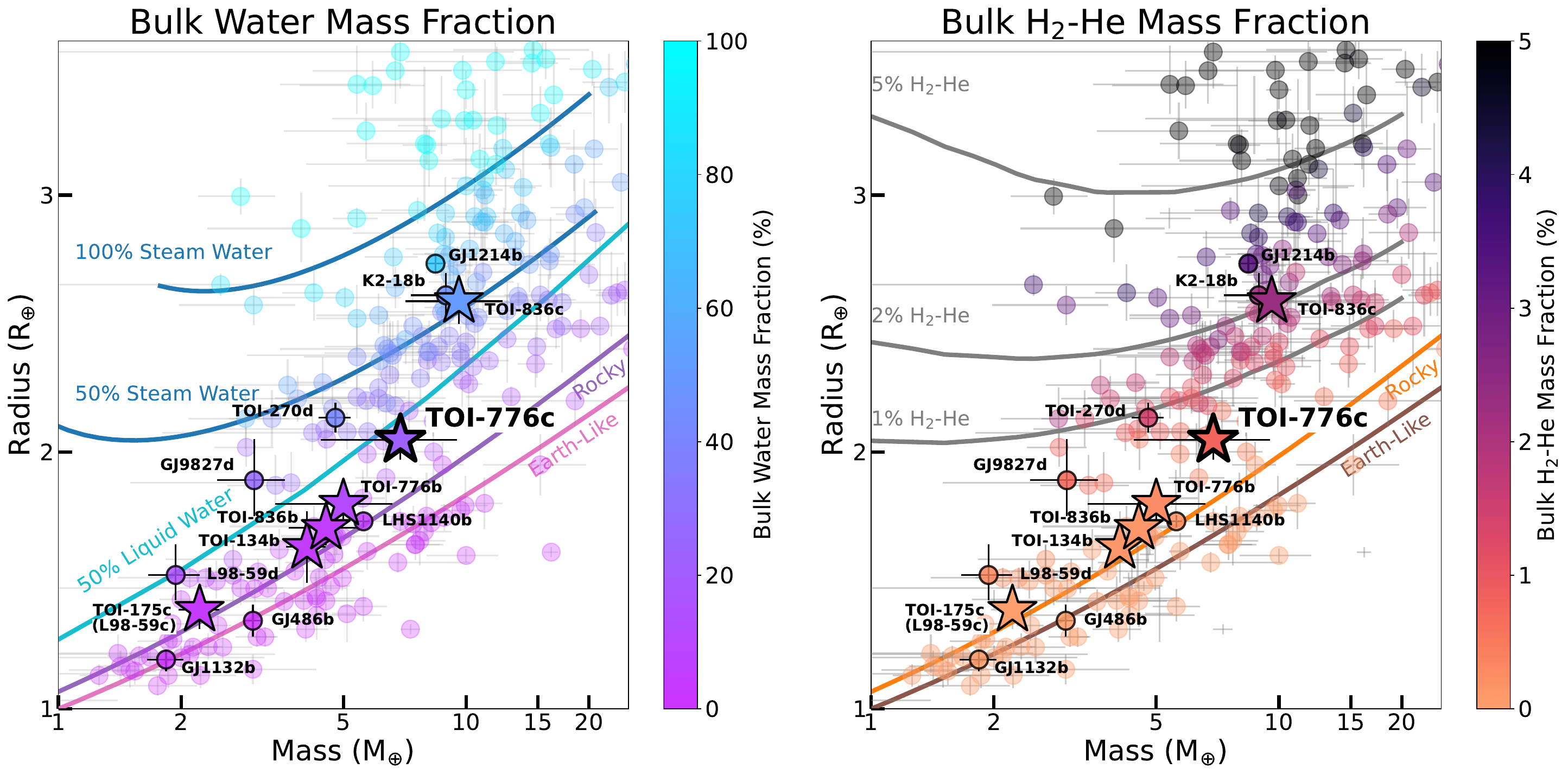}
    \caption{Masses and radii of well-constrained small planets ($ 0.2\leq M_p \leq 30$~M$_{\oplus}$). This catalog is a subset of the NASA Exoplanet Archive showing planets with consistently measured masses (RV or TTV) and measured radii (transit). The points are 
     color-coded by composition: On the left, as bulk water mass fraction \citep{Aguichine2021} and on the right, as bulk H$_2$/He mass fraction \citep{Lopez2014}. Composition lines are shown for an equilibrium temperature of 500~K; rocky and Earth-like predictions are from \cite{Zeng2016}. Bold and labeled circle symbols show planets with published JWST transmission spectra, and bold and labeled star symbols show other published COMPASS targets. TOI-776~c (the subject of this paper) is shown in larger font/symbol.}
    \label{fig:mr_both}
\end{figure*}

\begin{figure}
    \centering
    \includegraphics[width=0.5\textwidth]{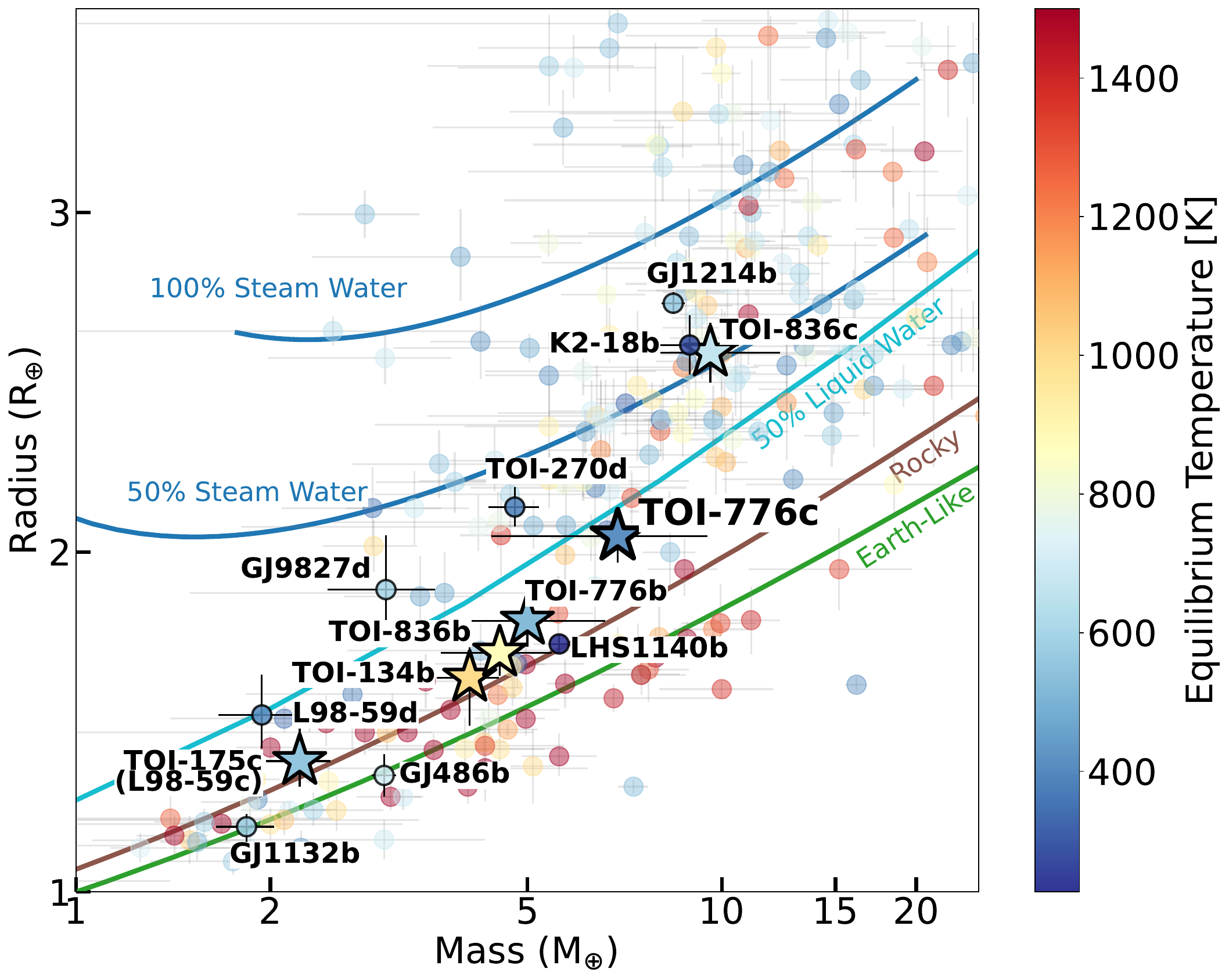}
    \caption{Similar to Figure~\ref{fig:mr_both}, but color-coded by equilibrium temperature (assuming albedo=0, full heat redistribution).}
    \label{fig:mr_teq}
\end{figure}

The TOI-776 system is an interesting benchmark for understanding planet formation around M dwarf stars. Both planets lie within the radius valley, a region of low planet occurrence in the period-radius space of small planets.
Recent work further exploring the stellar mass dependence of the radius valley \citep{Ho2024} suggests that, given the host star $M_* \sim 0.5$~M$_{\odot}$, TOI-776~b is on the lower radius edge of the valley and TOI-776~c is closer to the center of the valley. Planets in this regime are particularly mysterious as they do not fit neatly into either the bare rock or gas dwarf categories thought to be represented by the two occurrence rate peaks on either side of the radius valley.  
As described above, \cite{Luque2022Sci...377.1211L} put the TOI-776 planets in the category of ``water worlds''. To date, there are two other planets in this regime with published JWST transmission spectra: TOI-270~d \citep{2024arXiv240303325B}, which is similar to TOI-776~c in size, temperature, and host star mass (M3V vs. M1V) but is slightly less dense, and GJ 9827~d \citep{Piaulet-Ghorayeb2024}, which is similar in size but is substantially less dense, is hotter ($\sim675$~K), and orbits a more massive star (K7V). Figure \ref{fig:mr_both} shows where TOI-776~c falls in mass-radius space; the points are colored 
according to a bulk compositions, where the envelope is made of water (left) or H$_2$/He (right). For each planet, we infer the bulk composition based on the planet's mass, radius, equilibrium temperature and system's age using either the interior model of \cite{Aguichine2021} for steam worlds, which assumes a pure steam atmosphere and supercritical envelope on top of an Earth-like core, or the interior model of \cite{Lopez2014}, which assumes a solar metallicity atmosphere and envelope on top of an Earth-like core. The range of possible compositions of each planet is inferred by MCMC using the emcee sampler \citep{Foreman-Mackey2013}, and the color corresponds to the mode of that distribution. Details on the priors and likelihood, as well as examples of such runs, can be found in \cite{Wallack2024} and \cite{Alderson2024}. We also show the same mass and radius data colored by equilibrium temperature in Figure \ref{fig:mr_teq}. TOI-776~c stands out in density as compared to the currently published planets observed in transmission with JWST, and is especially conspicuous within the COMPASS sample, making it an interesting target for atmospheric characterization.

In \S \ref{sec:obs} we describe our JWST observation set-up and data quality and in \S\ref{sec:data_reduction} our data reduction techniques and light curve fitting methods. In \S\ref{sec:results} we present the resulting transmission spectrum of TOI-776~c, and in \S\ref{sec:discussion} discuss our interpretation and place the spectrum/planet in context with other studied small planet atmospheres. We summarize our findings in \S\ref{sec:summary}.

\begin{deluxetable}{ll}
\tablecaption{System Properties for TOI-776~c \label{table:system} }
\tablewidth{0pt}
\tablehead{\colhead{Property}&\colhead{Value}}
\startdata
\textbf{Star} & \\
$V$ (mag) & 11.536$\pm$0.041 \\
$J$ (mag) & 8.483$\pm$0.018 \\
$K_s$ (mag)  & 7.615$\pm$0.020 \\
T$_*$ (K)& 3275$\pm$60\\
R$_*$ (R$_\odot$)& 0.547$\pm$0.017\\
M$_*$ (M$_\odot$)& 0.542$^{+0.040}_{-0.039}$\\
log(g) & 4.8$\pm$0.1\\
$[$Fe/H$]_{*}$ (dex) & -0.21$\pm$0.08\\
$P_{\rm{rot}}$ (d) & 21.13$\pm$0.06 \\
Age (Gyr) & 6.1$^{+7.0}_{-15.1}$ \\
\hline
\textbf{Planet c} &  \\
Period (days)& 15.665323$^{+0.000075}_{-0.000070}$ \\
$K$ (m~s$^{-1}$) & 2.65$^{+0.99}_{-0.97}$ \\
Mass (M$_{\earth}$)&  6.9$^{+2.6}_{-2.5}$\\
Radius (R$_{\earth}$)& 2.047$^{+0.081}_{-0.078}$\\
$\rho$ (g~cm$^{-3}$) & 4.4$^{+1.8}_{-1.6}$ \\
$T_{\tt eq}$ (K)& 420$\pm$10\\
$e$& 0.089$^{+0.048}_{-0.054}$ \\
$\omega$ (\textdegree)&7$^{+58}_{-52}$\\
inclination (\textdegree)&89.49$^{+0.30}_{-0.20}$\\
a (AU)& 0.1001$^{+0.0022}_{-0.0024}$ \\
\enddata 
\tablenotetext{}{All values from \cite{Fridlund2024}. The $T_{\tt eq}$ values assume zero bond albedo.}
\end{deluxetable}

%%%%%%%%%%%%%%%%%%%%%%%%%%%%%%%%%%%%%%%%%%%%%%%%%%%%%%%%%%%
\section{Observations \label{sec:obs}}
%%%%%%%%%%%%%%%%%%%%%%%%%%%%%%%%%%%%%%%%%%%%%%%%%%%%%%%%%%%

\subsection{NIRSpec}

We observed two transits of TOI-776~c with JWST/NIRSpec, the first from 2023 May 11 20:09 UT to 2023 May 12 03:27 UT and the second on 2023 May 27 from 12:08-19:26 UT. We used the high-resolution ($R\sim$2700) G395H mode, which provides spectroscopy between 2.87--5.14\,$\mu$m across the NRS1 and NRS2 detectors (with a detector gap between 3.72--3.82\,$\mu$m). The observations were taken in the NIRSpec Bright Object Time Series (BOTS) mode using the SUB2048 subarray, the F290LP filter, the S1600A1 slit, and the NRSRAPID readout pattern. Each %6.38\,hr 
observation consisted of 3636 integrations (each 6.314 s) %(divided into three segments) 
with seven groups per integration and was designed to be centered on the transit events with sufficient out-of-transit baseline. Our observations resulted in $\sim1.62/2.74$ hours of pre-/post-transit baseline for both visits. The JWST data analyzed in this paper can be found in MAST: \dataset[10.17909/8tep-1j18]{http://dx.doi.org/10.17909/8tep-1j18}.

%%%%%%%%%%%%%%%%%%%%%%%%%%%%%%%%%%%%%%%%%%%%%%%%%%%%%%%%%%%
\section{Data Reduction} 
\label{sec:data_reduction}
%%%%%%%%%%%%%%%%%%%%%%%%%%%%%%%%%%%%%%%%%%%%%%%%%%%%%%%%%%%

We reduced the observations using two different pipelines, \texttt{Eureka!} \citep{2022JOSS....7.4503B}\footnote{https://github.com/kevin218/Eureka} and \texttt{Tiberius} \citep{Kirk2017,Kirk2021}, followed by two different light curve fitting schemes. 
Below we describe the details of each approach, and a comparison of the different choices/assumptions is summarized in Table \ref{table:redux_compare}.

\subsection{Eureka!}
\label{sec:eureka}
%Citations: \citep{2022JOSS....7.4503B}
%Edited from what Nicole wrote in the 776.02 paper (hopefully it's not too similar!):
We %first 
utilized %the open-source 
\texttt{Eureka!} 
%package 
to reduce the data in the same way as described in \cite{Alderson2024} and \cite{Wallack2024}. 
When reducing JWST observations, \texttt{Eureka!} acts as a wrapper for the initial two calibration stages of the STScI \texttt{jwst} pipeline. We used \texttt{Eureka!} v0.10 and \texttt{jwst} pipeline version 1.11.4 (context map jwst\_1225.pmap). We use the default Stage 1 and Stage 2 inputs for the \texttt{jwst} pipeline, except for %that we use 
a jump detection threshold of 15$\sigma$ in Stage 1. We also adopt 
the custom group-level background subtraction method that \texttt{Eureka!} includes to correct for the 1/$f$ noise known to be present in NIRSpec observations  \citep{2023Natur.614..664A}. The group-level background subtraction within \texttt{Eureka!} removes a column-by-column median from the trace-masked detector image. 

Stage 3 of \texttt{Eureka!} allows for a number of different reduction-specific choices in the calibration of extraction of the spectra. We optimize the choice of extraction apertures (full widths ranging between 8 -- 16 pixels) and background apertures (full widths ranging between 16 -- 22 pixels), polynomial order of an additional background subtraction, and sigma thresholds for optimal extraction outlier rejection by selecting the combination of those parameters that minimizes the median absolute deviation in flux as a function of time. The optimal combination of reduction parameters is given in Table~\ref{table:redux_compare}. We extract 30 pixel wide spectroscopic light curves and white light curves (42 bins between 2.863 and 3.714~$\mu$m for NRS1 and 64 bins between 3.820 and 5.082~$\mu$m for NRS2).

Using these light curves from the \texttt{Eureka!} reduction, we then deviate from the inbuilt \texttt{Eureka!} functionality for fitting the light curves and opt for a custom fitting routine (described in \citealt{Alderson2024,Wallack2024}) for added flexibility. For all light curves, we first iteratively trim three sigma outliers from a 50 point rolling median three times. We then use the  Markov chain Monte Carlo (MCMC) package \texttt{emcee} \citep{Foreman-Mackey2013} to simultaneously fit a transit and instrumental noise model to the light curves. For both the spectroscopic and white light curve fits, we initialize 3$\times$ the number of free parameters as walkers at the best fit values from a Levenberg-Marquardt minimization. For the white light curves, we fit for orbital inclination ($i$), the ratio of semi-major axis to the stellar radius ($a/R_{\star}$,) the mid-transit time ($T_{0}$), and the transit depth ($R_{p}/R_{\star}$) using \texttt{batman} \citep{Kreidberg2015} and holding the eccentricity ($e$), longitude of periastron ($\omega$), and period ($P$) fixed to the values in Table~\ref{table:system}. We fix the quadratic limb-darkening coefficients to the theoretical values computed with the model from the Set One of the MPS-ATLAS grid \citep{Kostogryz2022} closest to the stellar parameters in Table~\ref{table:system} using {\tt ExoTiC-LD} \citep{Grant2024}. We also test reproducing the Eureka! spectrum but with free limb darkening parameters. The resulting transmission spectrum (with limb darkening as a free parameter) has 4 ppm and 8 ppm worse precision in NRS1 and NRS2, respectively, and is almost identical to the fixed limb darkening case (agrees to well within $1\sigma$, $\lesssim$25 ppm). We move forward with analyzing the fixed limb-darkening spectrum. 

\begin{figure*}
    \centering
    \includegraphics[width=\textwidth]{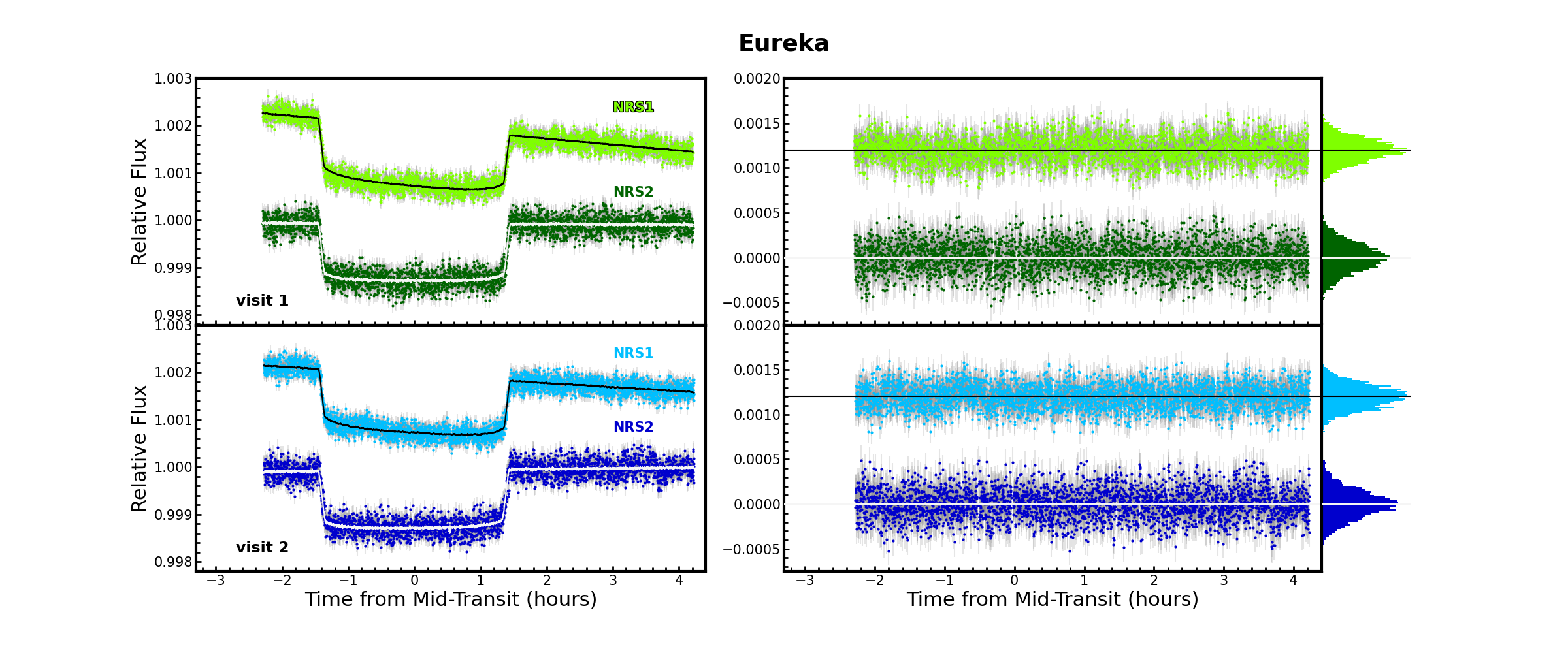}
    \caption{Left: The \texttt{Eureka!} white light curves of TOI-776 c from visit 1 (top, green) and visit 2 (bottom, blue) for each NIRSpec detector, offset for clarity. Error bars are shown in grey. Our best fit models, including systematics, are overplotted in black and light grey. Right: The residuals from the white light curve fits, offset for clarity. Black and grey lines at a residual value of zero are shown to guide the eye. The distribution of residual errors is shown on the far right. The time scale is purposefully set to match Figure~\ref{fig:tiberius_wlc} for ease of comparison.}
    \label{fig:eurkea_wlc}
\end{figure*}

\begin{figure}[h]
    \centering
    \includegraphics[width=0.49\textwidth]{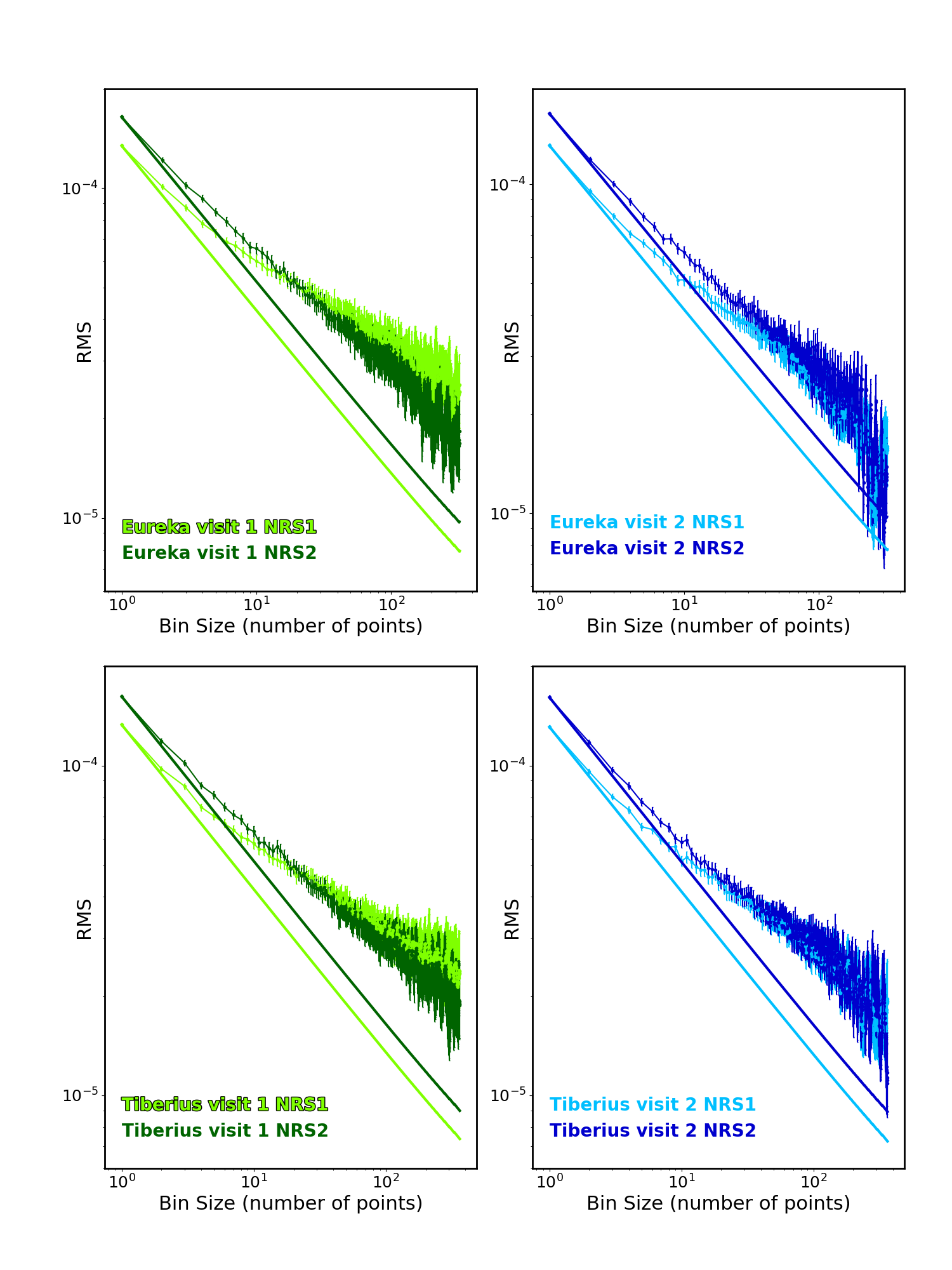}
    \caption{The white light curve Allan deviation plots for each detector and visit from the \texttt{Eureka!} (top) and \texttt{Tiberius} (bottom) reductions. Idealized (Gaussian, white) noise would fall along the solid line in each plot. Instead, we see that each detector has some level of correlated (red) noise, probably due to instrumental systematics that are unaccounted for in our models. These plots were made with the \texttt{mc3:}Multi-Core Markov-Chain Monte Carlo package \citep{Cubillos2017}.}
    \label{fig:allan_combined}
\end{figure}

\begin{figure}
   \centering
    \includegraphics[width=0.4\textwidth]{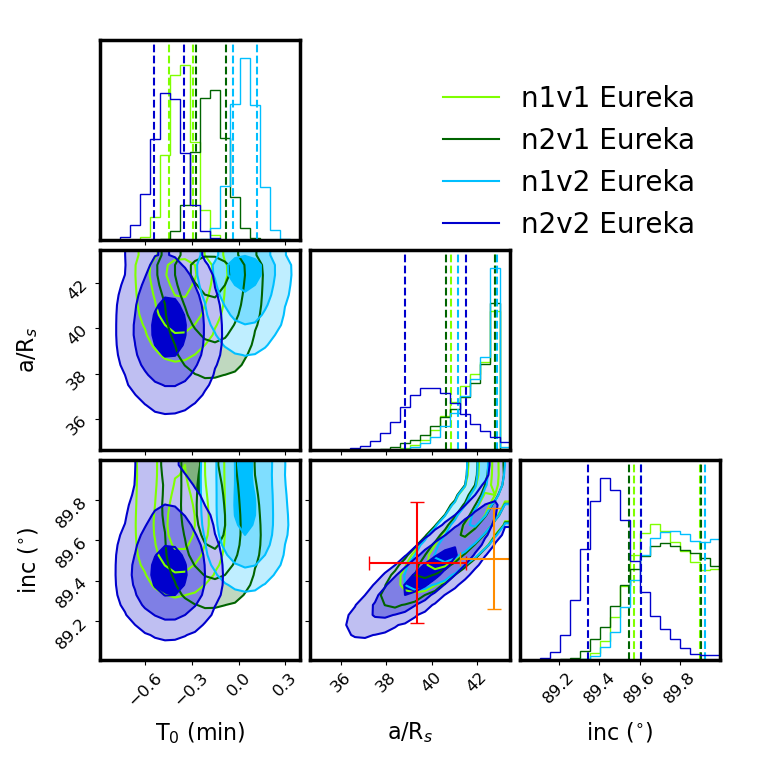}  \\  \vspace{-0.2cm} 
    \includegraphics[width=0.4\textwidth]{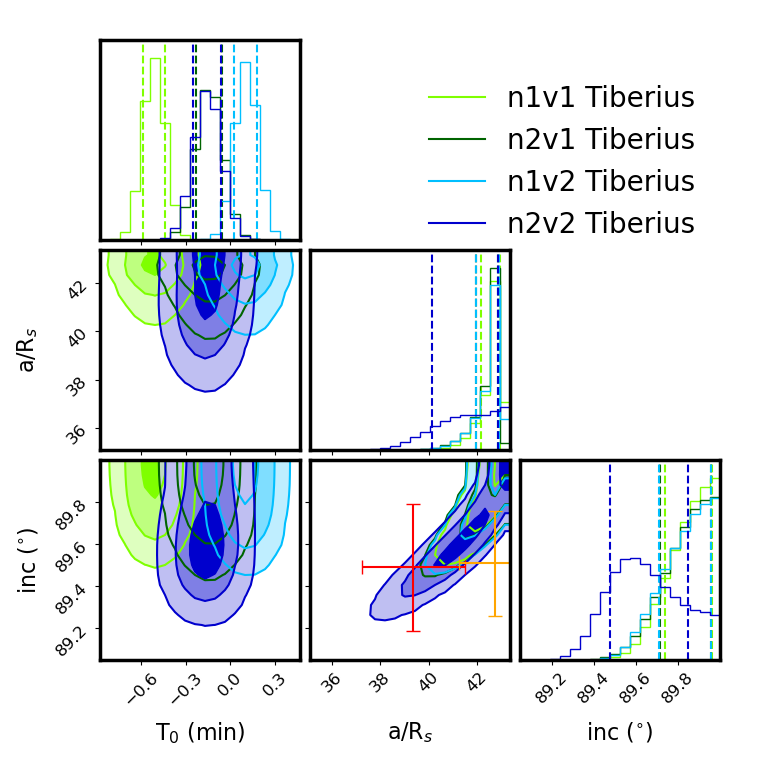}\\ \vspace{-0.2cm} 
    \includegraphics[width=0.4\textwidth]{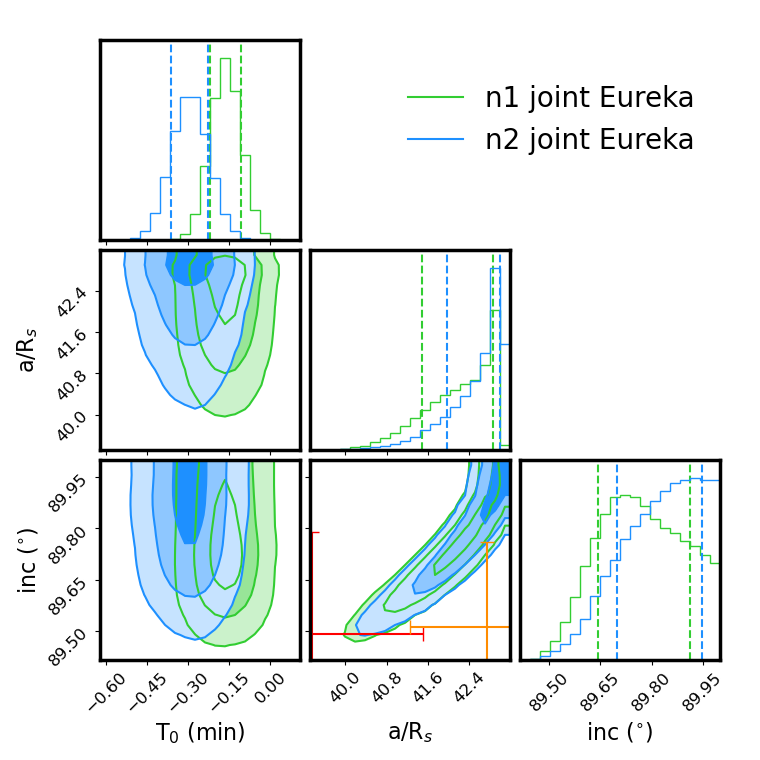}\\ \vspace{-0.13cm} 
    \caption{The posteriors for the time (T$_0$ in min from predicted mid-transit times of MBJD 60075.969 and 60091.634, 
    %60075.968699999794 and 60091.633999999794, 
    as in Table \ref{table:wlc_results}), $a/R_*$, and inclination from fitting the \texttt{Eureka!} (top) and \texttt{Tiberius} (middle) white light curves from each visit, and from the \texttt{Eureka!} joint fit of both visits (bottom). Contours show the 1, 2, and 3$\sigma$ confidence intervals, and the vertical dashed lines show the 16\%, 50\%, and 84\% quantile values. In red and orange we show the values of a/R$_*$ and inclination from \cite{Fridlund2024} and \cite{Luque2021}, respectively.}
    \label{fig:wlc_corner_plots}
\end{figure}

Our instrumental noise model, \textit{S}, is of the form
\begin{equation}
S= p_{1} + p_{2}\times T+ p_{3}\times X + p_{4}\times Y , 
\label{eq:1}
\end{equation}
where $p_{N}$ is a set of coefficients fitted for in our noise model, $T$ is the array of times, and $X$ and $Y$ are arrays of the positions of the trace (in the dispersion and cross-dispersion directions, respectively). We fit the white light curve from each visit and each detector separately. Then, to fit the individual spectroscopic light curves, we fix the $i$, $a/R_{\star}$, $T_{0}$ values to the white light curve values (given in Table~\ref{table:wlc_results}). 
For both the white light curves and the spectroscopic light curves, we fit for an additional per-point error term that is added in quadrature to the measured errors. We also trim the first 400 points (48.12 minutes) of each observation to remove any initial ramp that may bias the slope of the noise model. In addition to fitting the individual visits, we also jointly fit the two visits together (keeping the detectors separate), assuming the same astrophysical parameters in the same manner as described above. The joint fit parameters are also listed in Table~\ref{table:wlc_results}. The fitted white light curves and residuals are shown in Figure \ref{fig:eurkea_wlc}, the Allan deviation plots are shown in Figure~\ref{fig:allan_combined}, and the posteriors from the fits to the white light curves are shown in Figure~\ref{fig:wlc_corner_plots}. Figure \ref{fig:visit_spec} shows our \texttt{Eureka!} spectra for both the individual visits and the joint fit. The individual visit spectra have median uncertainties of 27 ppm for NRS1 and 47 ppm for NRS2 in our 30 pixel ($\sim0.02~\mu$m) bins, whereas the jointly fit spectrum has a median uncertainty of 19 and 33 ppm for NRS1 and NRS2, respectively.

\begin{figure*}
    \centering
    \includegraphics[width=\textwidth]{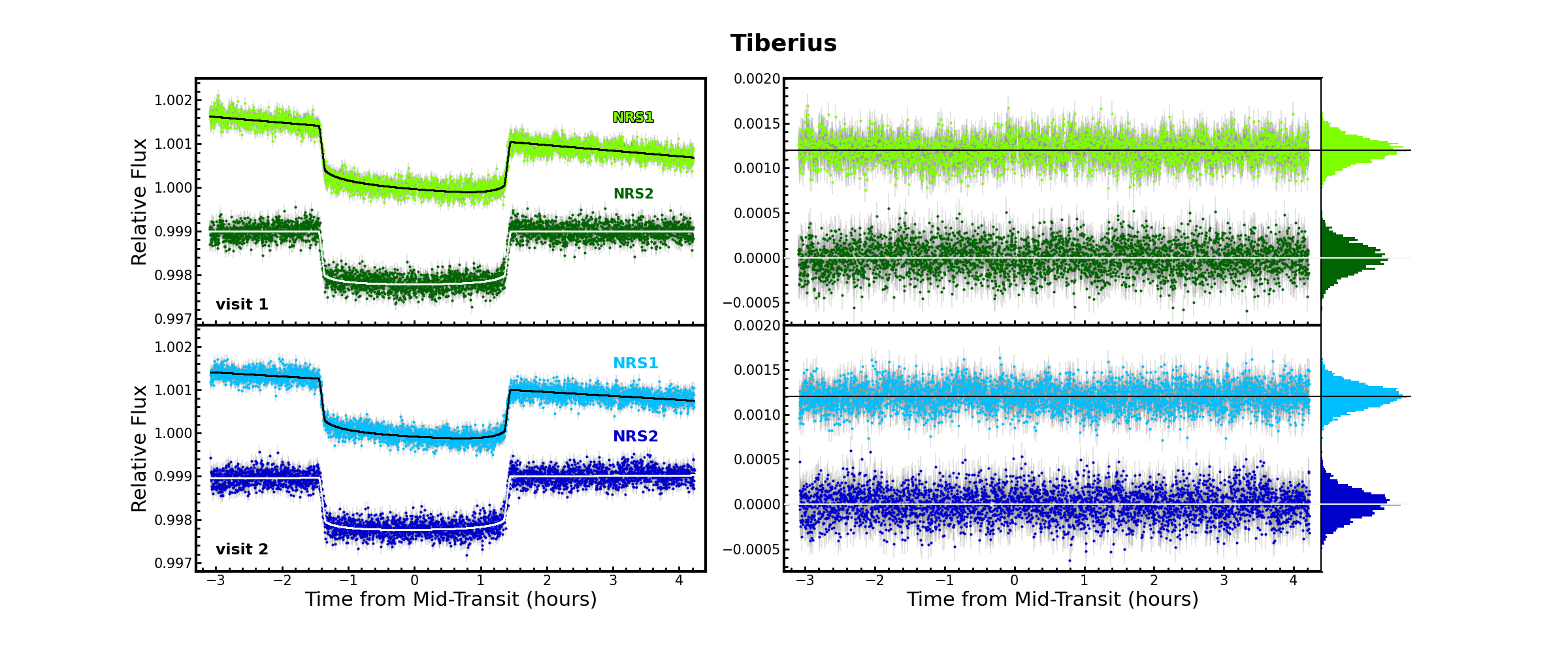}
    \caption{Similar to Figure~\ref{fig:eurkea_wlc}, except for the \texttt{Tiberius} reduction.}
    \label{fig:tiberius_wlc}
\end{figure*}

\subsection{Tiberius}
\label{sec:tiberius}

We also reduced the data using the \texttt{Tiberius} pipeline \citep[v1.0.0;][]{Kirk2017,Kirk2021}, which has been successfully used in a number of JWST studies \citep[e.g.,][]{2023Natur.614..649J,2023Natur.614..664A,2023ApJ...948L..11M,Kirk2024}. Our analysis of the TOI-776~c data begins by processing the \texttt{uncal.fits} files through stage 1 of the \texttt{jwst} pipeline (version 1.8.2). However, we apply our own 1/$f$ correction (group-level background subtraction), which involves subtracting the median of every pixel column after masking 22 pixels centered on the curved stellar traces. We then proceed to run spectral extraction on the \texttt{gainscalestep.fits} files. We also skip the \texttt{jump} step in favor of our own outlier detection. For this, we calculated the running median for every pixel's time series with a sliding box of three pixels and subtracted this running median from every pixel's time series. We then used the residuals to identify $5\sigma$ outliers and replaced these outliers with the running median's value. We then proceed to run our spectral extraction on the outlier cleaned integrations. We begin by locating the center of the stellar spectra trace by fitting a Gaussian to each column in the cross-dispersion direction, and then fit these centers with 
a quartic polynomial and perform standard aperture photometry with an aperture full width of 8 pixels, after removing the background at every column on the detector. The background is calculated from the median of the column rows after masking 7 pixels on either side of the 8 pixel aperture. Due to the curved nature of the stellar trace, the total number of background pixels is always $\geq 10$ but varies across the detector, depending on whether the 7 masked pixels include pixels that fall off the edge of the detector.

To obtain the wavelength array, we run the \texttt{assign\_wcs} and \texttt{extract\_2d} steps of the \texttt{jwst} pipeline. We then proceed to make our white light curves and spectroscopic light curves. Our white light curves are integrated over $2.73-3.72$\,$\mu$m for NRS1 and $3.82-5.18$\,$\mu$m for NRS2. For the spectroscopic light curves, we integrate over bins of 30 pixels ($\sim 0.02$\,$\mu$m) covering the same wavelength range as the white light curves. This results in 42 spectroscopic light curves for NRS1 and 64 for NRS2.

The models we fit to our light curves are comprised of an analytic (\texttt{batman}, \citealp{Kreidberg2015}) transit light curve multiplied by a linear polynomial. The free parameters in the fits to our white light curves are the scaled planetary radius ($R_P/R_\star$), the scaled semi-major axis ($a/R_\star$), the planet's orbital inclination ($i$), the time of mid-transit ($T_0$) and the two parameters of our linear polynomial. We hold the period ($P$), eccentricity ($e$) and longitude of periastron ($\omega$) fixed to the values of \cite{Luque2021} (15.6653\,d, 0.04 and -11$^{\circ}$, respectively) and place a uniform prior on $i$ of $\leq 90^{\circ}$. We find that without this prior, the strong degeneracy between $a/R_\star$ and $i$ leads to poor fits to the ingress and egress data. Including this prior, which is equivalent to putting a prior on the impact parameter $>0$ as done by \cite{Luque2021}, improves the fits to the ingress and egress data but ultimately leads to a much less than $1 \sigma$ difference in the transmission spectrum. The quadratic limb darkening coefficients are calculated using the 3D Stagger grid \citep{Magic2015} via \texttt{ExoTiC-LD} \citep{Grant2024} with the stellar parameters of \cite{Luque2021} ($\mathrm{T_{eff}} =  3709$\,K, [Fe/H] = -0.20, $\log g_\star = 4.727$\,cgs). In all of our fits, we hold the limb darkening coefficients fixed to the model values. We also ran a full end-to-end reduction with free limb darkening and found this led to a transmission spectrum that agreed with the fixed limb darkening spectrum to within $1 \sigma$ but with 8\,\% worse precision. 

Our white light curve fits, shown in Figure~\ref{fig:tiberius_wlc}, have six free parameters per visit and detector, which are fitted independently. Prior to fitting the light curves, we first clip $4\sigma$ outliers in the light curves using a running median. We run a Markov Chain Monte Carlo (via \texttt{emcee}) to infer our best-fitting values. We run two iterations, each with 120 walkers and 50000 steps. The first iteration is used to rescale the photometric uncertainties to achieve $\chi^2_\nu = 1$. The second iteration is run with the rescaled uncertainties. Our final parameter values and uncertainties are taken from the final 37500 steps of the second iteration. We list these in Table \ref{table:wlc_results}, with posteriors shown in Figure~\ref{fig:wlc_corner_plots}, and note that we have inflated our inclination upper uncertainty to include $90^\circ$ given our prior. Due to the correlation between $i$ and $a/R_\star$, we have also inflated our upper uncertainty on $a/R_\star$. In Figure~\ref{fig:allan_combined} we show the corresponding Allan deviation plots for our white light curve fits, which illustrate how our residuals compare to idealized (Gaussian, white) noise.

For the fitting of our spectroscopic light curves we set $a/R_\star$, $i$ and $T_0$ to fixed values from the mean-weighted values from the white light curve fits. This procedure results in three free parameters per spectroscopic light curve: $R_p/R_*$ plus the two coefficients of the linear polynomial. Again we clip $4\sigma$ outliers from the time-series. Instead of performing MCMC parameter estimation as was done for the white light curves, we run two iterations of a Levenberg-Marquadt algorithm for our spectroscopic light curves. The first iteration is used to rescale our photometric uncertainties and the second to obtain our final parameter values (the transmission spectrum). Our \texttt{Tiberius} transmission spectra have median uncertainties of 24 ppm for NRS1 and 42 ppm for NRS2 in our 30 pixel ($\sim0.02~\mu$m) bins; the weighted average spectrum has medium uncertainties of 17 and 30 ppm for NRS1 and NRS2, respectively.

\begin{figure}
   \centering
    \includegraphics[width=0.45\textwidth]{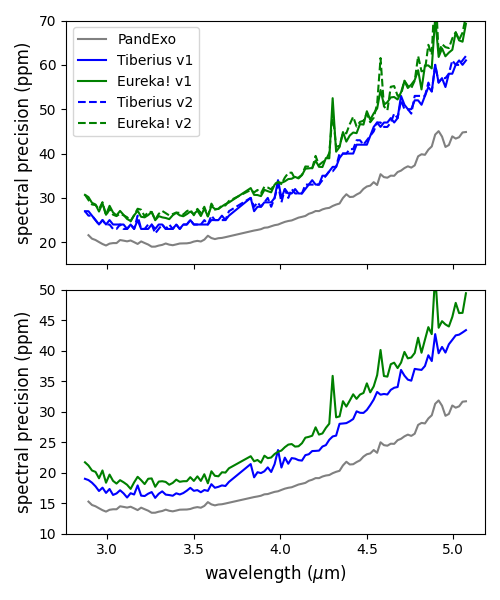}    
    \caption{Comparison of measured spectral precision versus the predicted \texttt{PandExo} precision for both \texttt{Eureka!} and \texttt{Tiberius} visit 1 and visit 2 (top panel), and joint visit fits (bottom panel).}
    \label{fig:pandexo}

\end{figure}

\begin{figure*}
    \centering
        \includegraphics[width=0.496\textwidth]{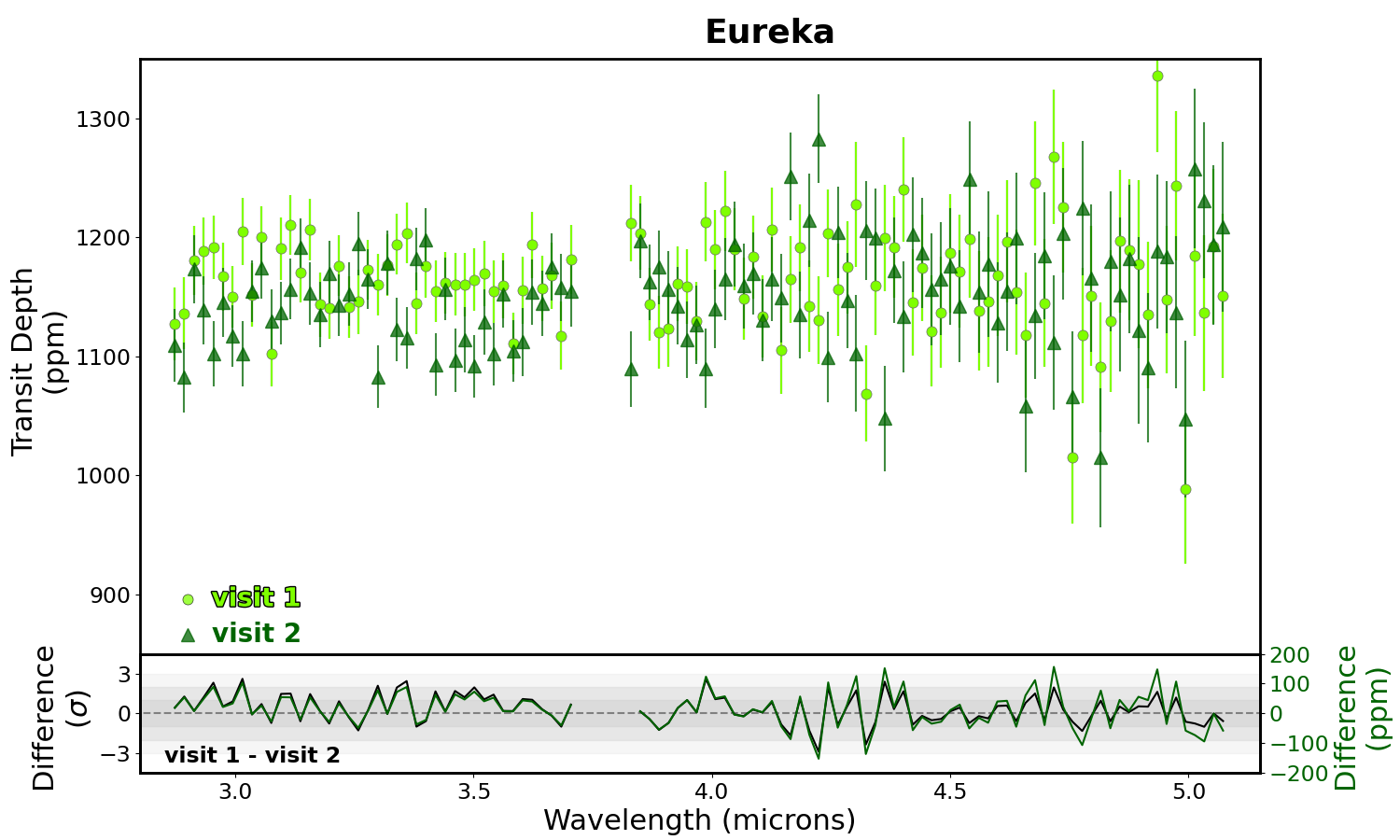}
    \includegraphics[width=0.496\textwidth]{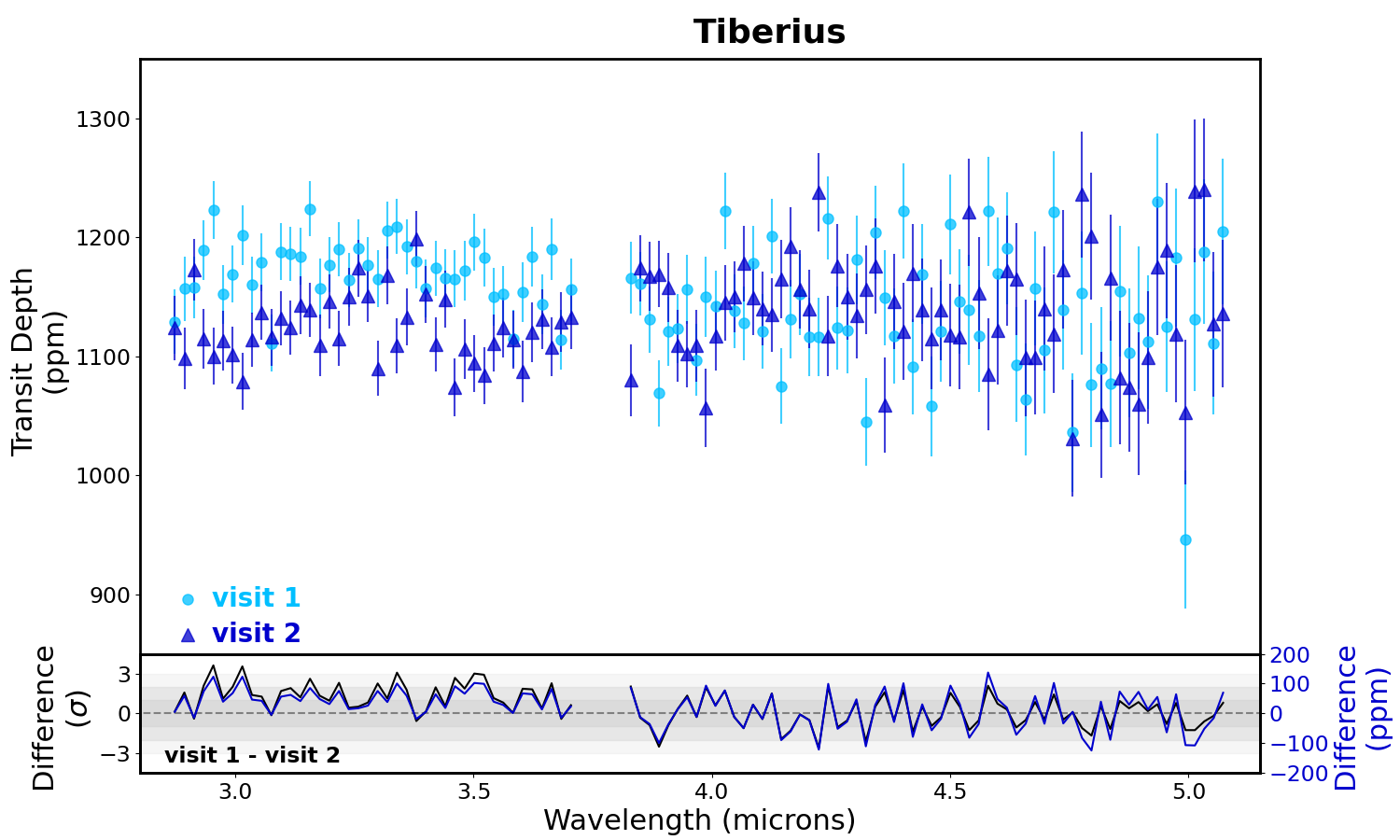}
    \includegraphics[width=\textwidth]{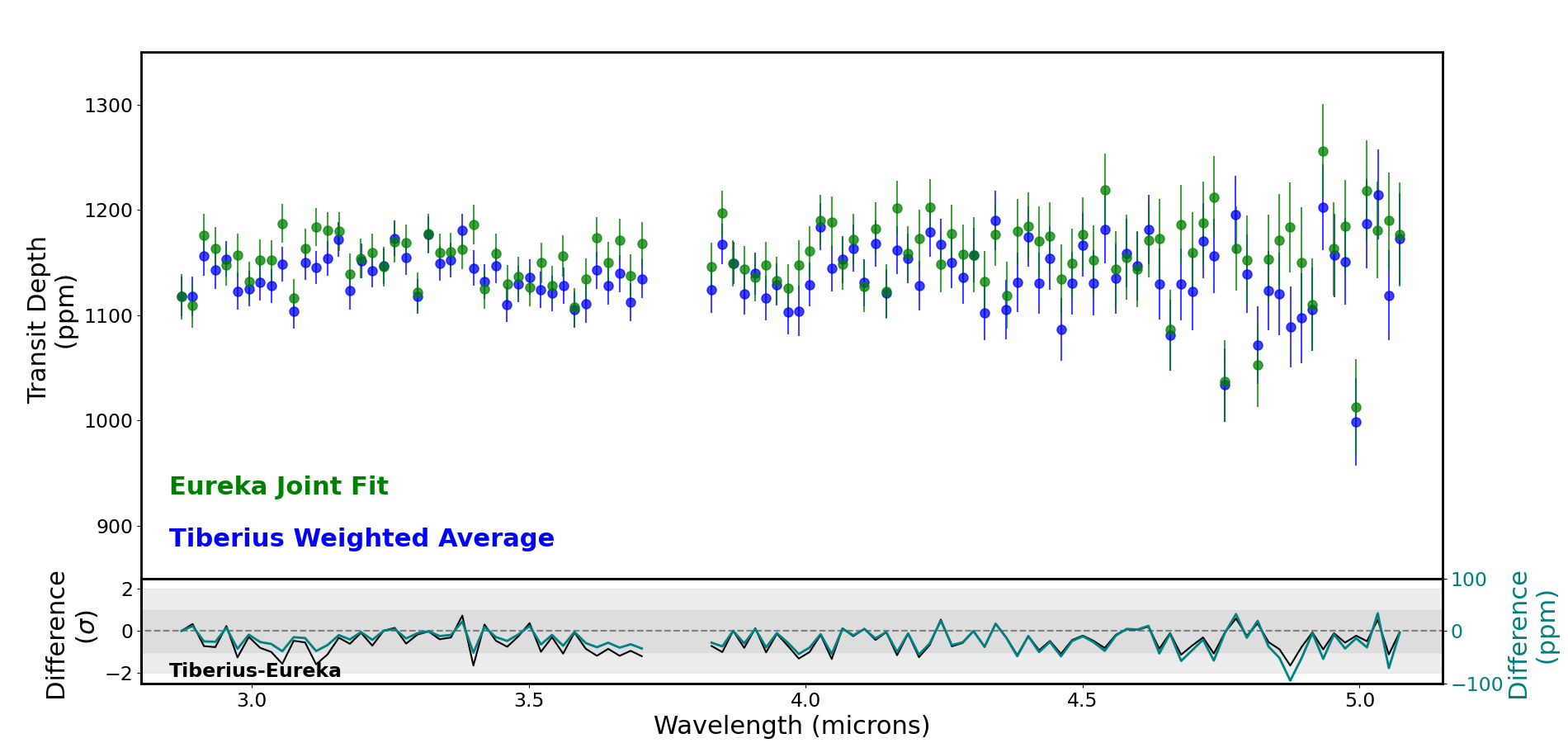}
    \caption{Top row: Spectra of TOI-776 c from each visit (light colors and circles vs. dark colors and triangles) and reduction pipeline (\texttt{Eureka!} on the left vs. and \texttt{Tiberius} on the right), in 30 pixel bins. The y-scale is the same in both plots. The bottom panel of each plot shows the differences between visits in units of sigma uncertainty (left axis) and ppm (right axis). Bottom row: Combined-visit spectra of TOI-776 c from each reduction pipeline, in 30 pixel bins. In green we show the \texttt{Eureka!} jointly fit spectrum (as described in \S\ref{sec:eureka}) and in blue the \texttt{Tiberius} weighted average spectrum (as described in \S\ref{sec:tiberius}). The bottom panel shows the differences between the two reductions in units of sigma uncertainty (left axis) and ppm (right axis). 
    }
    \label{fig:visit_spec}
\end{figure*}

In Figure~\ref{fig:visit_spec} we show the spectra resulting from the \texttt{Tiberius} reduction, both the individual visits and the weighted average. 

\subsection{Further Tests}
\label{sec:gp}

Given that it appears both reductions still contain some residual correlated noise (e.g., Figure~\ref{fig:allan_combined}), we also tested adding a Gaussian process (GP) component to the white light curve fits for both reductions, to represent the correlated noise. We used a simple harmonic oscillator kernel \citep{Foreman-Mackey2017} with the quality factor set to the critical value of $1/ \sqrt(2)$ and constraining the timescale parameters to be between 1/2 and 1/20$^{th}$ the total duration of the observation. We added this GP to the \texttt{Tiberius} fit (which also includes a linear term with time), as well as the \texttt{Eureka!} fit (both with and without the systematics model that includes the x/y position shift). We found that the resulting posteriors for the Visit 2 NRS2 $a/R_*$ and inclination were still discrepant from the other distributions, although there was better agreement between $T_0$ values between Visits 1 and 2 for NRS1. However, when we proceeded to use the median values from these fits including the GP component as input into the spectroscopic light curve fitting, we found no significant difference to the resulting transmission spectra. Thus we proceed with our results that do not include a GP component to the white light curve fit.

It also appears that at least the \texttt{Tiberius} NRS1 spectra are offset between visit 1 and visit 2. Comparing the visit 1 and visit 2 flux-calibrated stellar spectra (out of transit) does not reveal large changes between visits, but we performed the following test to investigate whether visit-to-visit changes could impact our spectra. First we took the visit 1:visit 2 ratio of the not-flux-calibrated stellar spectra extracted from \texttt{Tiberius}. We fit a polynomial to this ratio, and divided the visit 1 data by empirical correction, then refit the light curves to derive a transmission spectrum. The ``empirically-corrected'' \texttt{Tiberius} visit 1 transmission is virtually indistinguishable from that without any correction. Thus we proceed with our results that do not include an empirical correction, and account for the visit offsets in our modeling analysis below.

\begin{deluxetable*}{ccc}
\tabletypesize{\footnotesize}
\tablecolumns{3}
%\tablewidth{0pt}
\tablecaption{Data Reduction Choices/Assumptions \label{table:redux_compare}}
\tablehead{
\colhead{} & \colhead{\texttt{Eureka!}} & \colhead{\texttt{Tiberius}}}
\startdata
\texttt{jwst} pipeline version & 1.11.4 & 1.8.2 \\
Group-level background subtraction (remove $1/f$ noise) & column-by-column median & column-by-column median \\
\hline
Outlier detection/sigma clipping & Full frame 10$\sigma$,  & Full frame 5$\sigma$, \\
                                 &  light curves 3$\sigma$ iterative rolling median & light curves 4$\sigma$ \\
\hline
Extraction & Optimal & Standard \\
\hline
Aperture full width (pixels)   & 8  & 8 \\
          & (except for Visit 2, NRS1, which was 10) &     \\    
\hline
Pixels used to  & 8 pixels from top  & 7 pixels away from the trace \\
construct background &  and bottom of detector & aperture on both sides \\
\hline
Integration-level background subtraction &  column-by-column median    &  column-by-column median\\     
\hline
Spectral range extracted ($\mu$m)      &   2.863-3.714 (NRS1) & 2.730-3.719 (NRS2) \\
                                       &   3.820-5.082 (NRS2) & 3.822-5.175 (NRS2) \\
\hline
Spectroscopic bin width ($\mu$m, pixels)      &   0.02, 30 & 0.02, 30 \\
\hline
\# of trimmed integrations at start$^{*}$            & 400     & 0 \\
\hline
White-light curve (WLC) model &   linear polynomial & linear polynomial \\
                        &   (p1$+$p2$\times t$ + p3$\times X$ + p4$\times Y$)$^{\dagger}$ &  p1$+$p2$\times t$ \\                       
\hline
Spectroscopic light curve (SCL) model &   linear polynomial & linear polynomial \\
                        &   (p1$+$p2$\times t$ + p3$\times X$ + p4$\times Y$)$^{\dagger}$ & p1$+$p2$\times t$ \\                       
\hline
Fitting process & MCMC (WLC), MCMC(SLC),  & MCMC (WLC), LM (SLC), \\
                & $i<90^{\circ}$ prior &  $i<90^{\circ}$ prior \\
\hline
Limb-darkening & quadratic, fixed $u_1$, fixed $u_2$ & quadratic, fixed $u_1$, fixed $u_2$; \\
                & mps1 grid  &  3D Stagger grid used \\
\enddata
\tablecomments{$^{*}$No integrations were trimmed at the end in either reduction. $^{\dagger}$Here, $t$ is time, $X$ is x-position of the trace on the detector (the wavelength direction), and $Y$ is y-position of the trace on the detector; the other values are fitted coefficients.}
\end{deluxetable*}

\begin{table*}[htbp]
\setlength{\tabcolsep}{1pt}
%\renewcommand{\arraystretch}{1.0}
%\begin{landscape}
  \centering
\caption{Results of White Light Curve Fitting \label{table:wlc_results}}
%\begin{sideways}
\begin{footnotesize}
    \begin{tabular}{|c|c|c|c|c|c|c|c|c|}
    \hline
%    \makesavenoteenv{table*}
     & & & WLC    & SLC  & R$_{p}$/R$_{*}$ & T$_0$$^\dag$ & a/R$_*$ & i ($^\circ$) \\
    & & &  RMS (ppm) & RMS (ppm) &  &  &  &  \\
   \hline
    \texttt{Eureka!} & Visit 1 & NRS1 & 135& 675 & $0.0342508 \pm 9.3\times 10^{-5}$ & $-0.00026 \pm 5.5\times 10^{-5}$& 42.05$\pm$0.92 & 89.72$\pm$0.15  \\
                    &          & NRS2 &164 &1173 & 0.034019$\pm 1.0\times 10^{-4}$ & $-0.00012 \pm 6.5 \times 10^{-5}$ & 42.08$\pm$1.09 & 89.73$\pm$0.16  \\
                    &  Visit 2 & NRS1 & 131 &686& 0.033796$\pm 9.2\times 10^{-5}$ & $0.00003 \pm 5.5 \times 10^{-5}$ &42.33$\pm$0.87 & 89.77$\pm$0.14  \\
                    &          & NRS2 &164 & 1177 & 0.034249$\pm 1.1 \times 10^{-4}$ & $-0.00031 \pm 6.7 \times 10^{-5}$ &40.10$\pm$1.34 & 89.46$\pm$0.14  \\
                    & Joint Fit & NRS1 & 135 (visit 1), & 676 (visit 1), & 0.034009$\pm 6.5\times10^{-5}$ &  $-0.00011 \pm 3.9 \times 10^{-5}$ &42.31$\pm$0.66 & 89.77$\pm$0.12  \\
                      &  &  &  132 (visit 2) &   686 (visit 2) &  &   &&  \\

                    &           & NRS2 & 165  & 1173 (visit 1), &0.033879$\pm 7.5\times 10^{-5}$ & $-0.00020 \pm 4.8 \times 10^{-5}$ &42.72$\pm$0.56 & 89.843$\pm$0.11  \\
    &           &  &  (both visits) &  1177 (visit 2) &  &  & &  \\
     \hline
    \texttt{Tiberius} & Visit 1 & NRS1 & 133& 706& 0.034251$^{+7.3 \times 10^{-5}}_{-7.0 \times 10^{-5}}$ & $-0.000358 \pm 5.3 \times10^{-5}$ &42.77$^{+0.37}_{-0.60}$ & 89.87$\pm$0.13  \\
                    &          & NRS2 & 162& 1213&  0.033754$^{+8.8\times10^{-5}}_{-8.6 \times 10^{-5}}$ & $-0.00010 \pm 6.2 \times10^{-5}$ & 42.66$^{+0.46}_{-0.68}$ & 89.86$\pm$0.14  \\
                    &  Visit 2 & NRS1 & 131& 703&  0.033517$^{+7.7 \times 10^{-5}}_{-7.2 \times 10^{-5}}$ &  $0.00007 \pm 5.4 \times10^{-5}$ &42.70$^{+0.44}_{-0.73}$ & 89.86$\pm$0.14 \\
                    &          & NRS2 & 161& 1236&  0.033888$^{+1.13 \times 10^{-4}}_{-1.03 \times 10^{-4}}$ & $-0.00011 \pm 6.5 \times10^{-5}$ &41.63$^{+1.74}_{-1.48}$ & 89.64$^{+0.36}_{-0.16}$  \\
                    & Weighted  &  & & & 0.033863$\pm 4.1 \times 10^{-5}$ & $-0.00013 \pm 2.9 \times 10^{-5}$ &42.77$\pm$0.31 & 89.84$\pm$0.08 \\
                     &  Mean &  & & &  & & &  \\
     \hline
     \multicolumn{9}{l}{\vtop{\hbox{\strut $^\dag$Time from respective visit 1 and visit 2 expected mid-transit times MBJD 60075.968699999794 and 60091.633999999794.}}}
    \end{tabular}
    \end{footnotesize}

\end{table*}

%%%%%%%%%%%%%%%%%%%%%%%%%%%%%%%%%%%%%%%%%%%%%%%%%%%%%%%%%%%
\section{Interpretation of Planet's Transmission Spectrum} 
\label{sec:results}
%%%%%%%%%%%%%%%%%%%%%%%%%%%%%%%%%%%%%%%%%%%%%%%%%%%%%%%%%%%
In order to interpret the transmission spectra of TOI-776 c, we follow the procedures used in \cite{Wallack2024} and \cite{Alderson2024}. Namely, we begin with a nonphysical modeling approach to interpret the spectra on a visit-by-visit and reduction-by-reduction basis. Non-physical models are %important 
useful for understanding the basic structure and significance of features in the data %feature importance of the data, 
prior to comparison with physical models. For example, a non-physical model offers an avenue to statistically quantify whether or not there are non-zero slopes, instrument offsets, and/or Gaussian-like features in the spectra. After conducting our non-physical model interpretation, we use a physical model (\texttt{PICASO}, \citealt{2019ApJ...878...70B}) to understand how well the data can rule out certain atmospheric scenarios. For spectra that are largely void of features, the strategy employed in many analyses (e.g., \citetalias{2023arXiv230104191L,2023ApJ...948L..11M}; \citealt{Wallack2024}) has been to utilize atmospheric metallicity as a proxy for mean molecular weight of the atmosphere. We follow this approach %trend 
to utilize metallicity, and also conduct simple two-component atmosphere models (H$_2$O+H$_2$/He, CO$_2$+H$_2$/He) to rule out parameter space in mean molecular weight. We discuss caveats to all of these approaches below. We also aim to explore how aerosols (clouds and/or hazes) would hinder the inference of the limit placed on atmospheric mean molecular weight. Ultimately, these physically motivated models provide limits on mean molecular weight and aerosol pressure level (via an opaque pressure level). Below we describe these two analyses. 

\subsection{Feature Detection with Non-Physical Models}
We test four different nonphysical models: 1) a zero-sloped line (1 free parameter), 2) a zero-sloped line with a step function offset to account for baseline differences in NRS1 and NRS2 (2 free parameters), 3) a nonzero sloped line (2 free parameters),  4) a Gaussian feature in the NRS1 wavelength range (5 free parameters), and 5) a Gaussian feature in the NRS2 wavelength range (5 free parameters). The five free parameters for the Gaussian models include: transit baseline, feature amplitude, feature width, feature center, and the NRS1/2 offset. Based on the results of the NRS1 Gaussian test, we perform an additional test to quantify the significance of a potential feature at $\sim3.3~\mu$m (4 free parameters). This model has one less free parameter than test \#4 because we fix the wavelength center of the Gaussian at 3.314~$\mu$m while fitting for an amplitude and Gaussian width. For each case we use the \texttt{UltraNest} \citep{ultranest2021JOSS....6.3001B} package, which uses a nested sampling approach to map the likelihood space of the desired parameters given a data set. We use this package to obtain the Bayesian evidence ($\ln$Z), a best fit model and corresponding reduced chi-squared ($\chi^2/N$), as well as the best-fit parameters and confidence intervals. We perform tests on each data reduction and each visit separately. We also perform these tests on the joint fit (in the case of \texttt{Eureka!}) and the mean weighted spectrum (\texttt{Tiberius}). We use the difference in Bayesian evidence to determine which nonphysical model is preferred given the data. 

\begin{figure*}
    \centering
    \includegraphics[width=\textwidth]{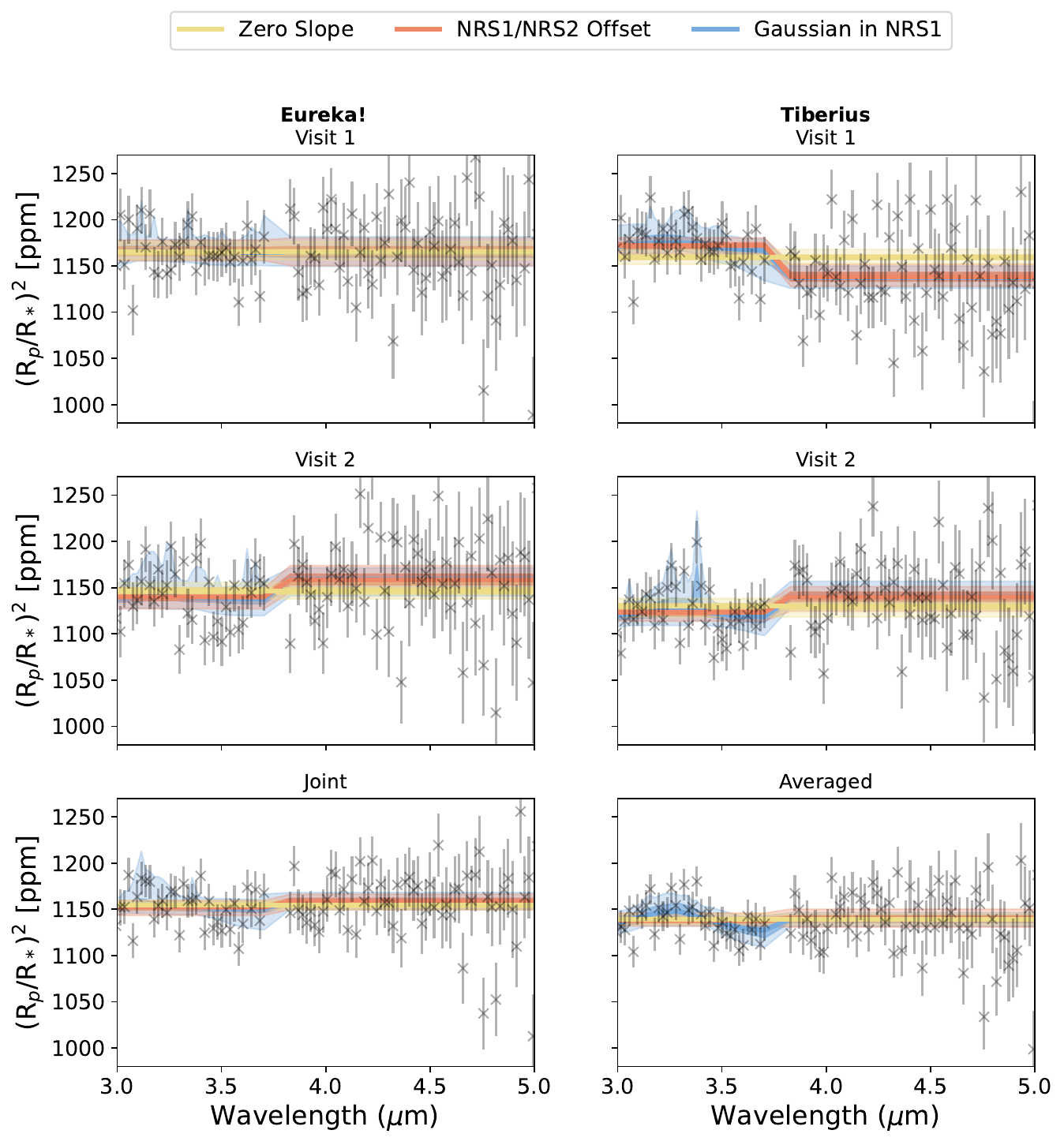}
    \caption{Here we show the transmission spectra detection for visit 1, visit 2 and the combined spectrum of TOI-776 c from our \texttt{Eureka!} (left) and \texttt{Tiberius} (right) reductions compared to the fitting results of the three  non-physical models with the highest Bayesian evidence. In all panels the data are grey x's with error bars, with the dark shaded region showing the 68-percentile region (1$\sigma$) and the light shaded region showing the 99-percentile region. Either a zero-sloped line or a zero-sloped line with an offset between NRS1/NRS2 is weakly preferred, at best, based on the relative Bayesian evidence (see Table \ref{tab:fits}).}
    \label{fig:nonphysical_feature_detection}
\end{figure*}

The results are shown numerically in Table \ref{tab:fits} and visually in Figure \ref{fig:nonphysical_feature_detection}. 
For the \texttt{Eureka!} data reductions, when comparing the Bayesian evidence of the zero-sloped model (null hypothesis) with that of the slope, and three Gaussian models, we find that the addition of higher model complexity is disfavored as it results in lower or equal values of lnZ. When comparing the zero-sloped model to the step function model, we disfavor the step model function for visit 1 and the joint spectrum, but weakly favor the addition of a step for visit 2 ($\Delta$lnZ=1).
The exact baseline of each of these reductions differs at the 1$\sigma$-level with transit depths between $\sim1150-1170$~ppm (see Table \ref{tab:fits} for exact values). $\chi^2/N$ for each of the best-fit models represents a good fit to the data, with values of 1.05-1.24 depending on which visit(s) are considered. 

For the \texttt{Tiberius} reductions, there is a strong preference for an offset between NRS1 and NRS2 for visit 1 ($\Delta\ln$Z=12). For visit 2, there is no preference for an offset, but there is a weak preference ($\Delta\ln$Z=1.3) for a Gaussian centered at $\sim$3.3~$\mu$m in NRS1 (see blue curve in Figure \ref{fig:nonphysical_feature_detection}). However, this weak preference disappears when visits 1 and 2 are averaged. No Gaussian feature is confidently detected in any of the cases for NRS2, given the large 1$\sigma$ constraint interval retrieved on the center wavelength. 

Given the strong interest in 3.3~$\mu$m, which is the location of a CH$_4$ ro-vibrational band, we take extra precaution to ensure the statistical significance of the \texttt{Tiberius} reduction is not preferred. We fix the position of the center wavelength at the CH$_4$ band and rerun the test on both \texttt{Eureka!} and \texttt{Tiberius} data sets. Fixing the center wavelength does not change the significance of the Gaussian test. Ultimately, we confidently rule out the presence of a feature in the place where CH$_4$ and CO$_2$ are expected by comparing the three three Gaussian tests with a zero-sloped model. 

\begin{table*}[]
\centering
\caption{Results of non-physical fits to Visit 1 \& 2 of both \texttt{Tiberius} and \texttt{Eureka!} data reductions as well as the combined final transmission spectra. In the last column we include the fit for the parameter listed in the ``Key Parameter'' column. They are: 1) the $(R_p/R_*)^2$ baseline intercept in ppm units for the zero-slope model, 2)  the offset between NRS1 and NRS2 in $(R_p/R_*)^2$ ppm units for the step function model, 3) the gradient of the slope (ppm/$\mu$m)  for the slope case, and 4)  the median wavelength in $\mu$m for the NRS1 and NRS2 Gaussian cases.}
\begin{tabular}{l|cccc}
                    & \multicolumn{4}{c}{\textbf{Eureka! (v1/v2/joint)}}           \\ \hline
\textbf{Model Type} &  $\ln$Z                & $\chi^2/N$ & Key Parameter & Parameter Fit \\ \hline
Zero Slope &  -60/-77/-70 & 1.05/1.37/1.24 & Baseline Intercept [ppm] & 1166$\pm$3/ 1147$\pm$3/ 1154$\pm$2  \\
Step Function  & -63/-76/-72 & 1.05/1.31/1.22   & NRS1/NRS2 Offset [ppm] & 0$\pm$7/ 19$\pm$7/ 6$\pm$5   \\
Slope & -63/-77/-73 & 1.05/1.31/1.23  & Slope Gradient [ppm/$\mu$m] & -2$\pm$6/15$\pm$6/4$\pm$4  \\
Gaussian NRS1 &  -64/-77/-72 & 1.05/1.29/1.21   & $\lambda_0$ [$\mu$m]  & 3.3$\pm$0.2/3.2$\pm$0.2/3.2$\pm$0.2 \\ 
Gaussian NRS2 &  -64.0/-77.0/-73.0 & 1.05/1.3/1.22 & $\lambda_0$ [$\mu$m] & 4.5$_{-0.5}^{+0.5}$/4.4$_{-0.3}^{+0.6}$/4.5$_{-0.5}^{+0.5}$ \\
Gaussian 3.314$\mu$m &  -64/-77/-72 & 1.05/1.3/1.22  & &  \\ \hline \hline

                    & \multicolumn{4}{c}{\textbf{Tiberius (v1/v2/weighted)}}            \\ \hline 
\textbf{Model Type} &  $\ln$Z                & $\chi^2/N$ & Key Parameter & Parameter Fit \\ \hline
Zero Slope & -91/-71/-72 & 1.64/1.27/1.28 &  Baseline Intercept [ppm] & 1159$\pm$3/ 1130$\pm$3/ 1139$\pm$2   \\
Step Function  &  -79/-71/-76 & 1.35/1.2/1.28 & NRS1/NRS2 Offset [ppm] & -34$\pm$6/ 16$\pm$6/ 1$\pm$4   \\
Slope & -80/-72/-75 & 1.35/1.23/1.28 & Slope Gradient [ppm/$\mu$m] & -30$\pm$5/12$\pm$6/-1$\pm$4  \\
Gaussian NRS1 &  -78/-70/-73 & 1.32/1.17/1.16 & $\lambda_0$ [$\mu$m] & 3.3$\pm$0.2/3.3$\pm$0.1/3.3$\pm$0.1 \\
Gaussian NRS2 &  -79.0/-72.0/-76.0 & 1.34/1.19/1.28 & $\lambda_0$ [$\mu$m]  & 4.4$_{-0.4}^{+0.6}$/4.4$_{-0.3}^{+0.6}$/4.4$_{-0.3}^{+0.5}$ \\
Gaussian 3.314$\mu$m &    -78/-70/-73 & 1.31/1.14/1.16 & & \\
 \hline \hline

\end{tabular}
\label{tab:fits}
\end{table*}

\subsection{Atmospheric Constraints from Physical Models}
We leverage physical models using two different statistical techniques (Bayesian analysis and forward modeling) to understand what parameter space can be ruled out in mean molecular weight and opaque pressure level. Both methods utilize the open source spectral model, \texttt{PICASO} \citep{2019ApJ...878...70B}. \texttt{PICASO}'s transmission spectrum code is based on the methodology developed in \cite{brown2001ApJ...553.1006B}. It utilizes the opacities released under the Zenodo Resampled Opacities v2 database \citep{natasha_batalha_2022_6928501}. Of particular importance are the opacities of CH$_4$, H$_2$O, and CO$_2$, where we utilize line lists of \citet{Hargreaves2020}, \citep{pokazatel}, and \citep{huang2014reliable}, respectively. Since our non-physical analysis (\S4.1) demonstrates that we do not detect any strong spectral features, the specific line lists used will not affect the results. \texttt{PICASO} automatically includes opacity sources for: CH$_4$, CO, CO$_2$, H$_2$O, K, NH$_3$, Na, TiO, VO, although for this study the spectral contributions from Na, K, TiO, VO are not present. For completeness, the full table of line lists used by \texttt{PICASO} can be found in \cite{Mukherjee2023ApJ...942...71M}. All temperature-pressure profiles are parameterized with the methodology described in \citet{Guillot2010A&A...520A..27G} with the equilibrium temperature for TOI-776 c (420 K, assuming zero albedo, full heat redistribution). Given this temperature, we compute chemical equilibrium abundances using the open source package \texttt{Cantera} \citep{Goodwin2022} with thermodynamics by \citet{Wogan2023} and solar elemental abundances from \citet{Lodders2009}. We keep the C/O ratio fixed at Solar (=0.458), and vary the metallicity from 1$\times$Solar to 1000$\times$Solar. Note that we also explore C/O ratio = 0.5 and 1.5 and find that the limits placed on M/H, discussed in the results, do not depend on C/O ratio. This is because at the low temperatures expected on TOI-776 c CH$_4$ remains a dominant source of CO$_2$. Opaque pressure levels, representative of a cloud deck, are included as an optically thick ($\tau$=10) slab at a given pressure.

In our forward modeling approach we compute a grid of models with 20 logarithmically spaced metallicity values, and five opaque pressure levels between 1 and 10$^{-4}$~bars. For each grid point we compute the reduced chi-squared per data point ($\chi^2/N$), which can be converted to a p-value (using \texttt{scipy.stats.distributions.chi2}) and a $\sigma$-confidence level (\texttt{scipy.stats.norm.ppf}, which assumes that the errors are normally distributed). Before $\chi^2/N$ is computed, we correct the data for any existing offsets using best-fit parameters from the step function model shown in Table \ref{tab:fits}. This procedure has been used in several other observations of featureless spectra \cite[e.g.][]{Wallack2024,Alderson2024}.

In our Bayesian analysis, we again use the \texttt{UltraNest} package. Here we fit for three physical (metallicity, opaque pressure level, radius factor) and two data driven parameters (NRS1/NRS2 offset, and error inflation term). The error inflation term ($f$) is added in quadrature to the derived error from either \texttt{Eureka!} or \texttt{Tiberius} ($\sigma_\mathrm{E/T}$ ) via: 
\begin{equation}
s_i^2 = \sigma_\mathrm{E/T}^2 + (10^f)^2
% e**2 + (10**(logf))**2
%    loglikelihood = -0.5*np.sum((y-resulty)**2/sigma2 + np.log(2*np.pi*sigma2))
\end{equation}
The log-likelihood is then computed as: 
\begin{equation}
\ln \mathcal{L}(y|\textbf{x}) = -\frac{1}{2} \sum_{i=1}^{N} \left[ \frac{(y_i - F_i(\textbf{x}))^2}{s_i^2} + \ln(2\pi s_i^2) \right]
% e**2 + (10**(logf))**2
%    loglikelihood = -0.5*np.sum((y-resulty)**2/sigma2 + np.log(2*np.pi*sigma2))
\end{equation}
where $y_i$ represents the data, and $F_i(x)$ represents the forward transmission spectral model for each given state vector, $\textbf{x}$=[M/H, opaque pressure level, radius factor, NRS1/NRS2 offset]. Unlike the forward modeling approach, this methodology allows the detector offsets and error inflation to vary along with the physical parameters.

\begin{figure*}
    \centering
    \includegraphics[width=0.9\textwidth]{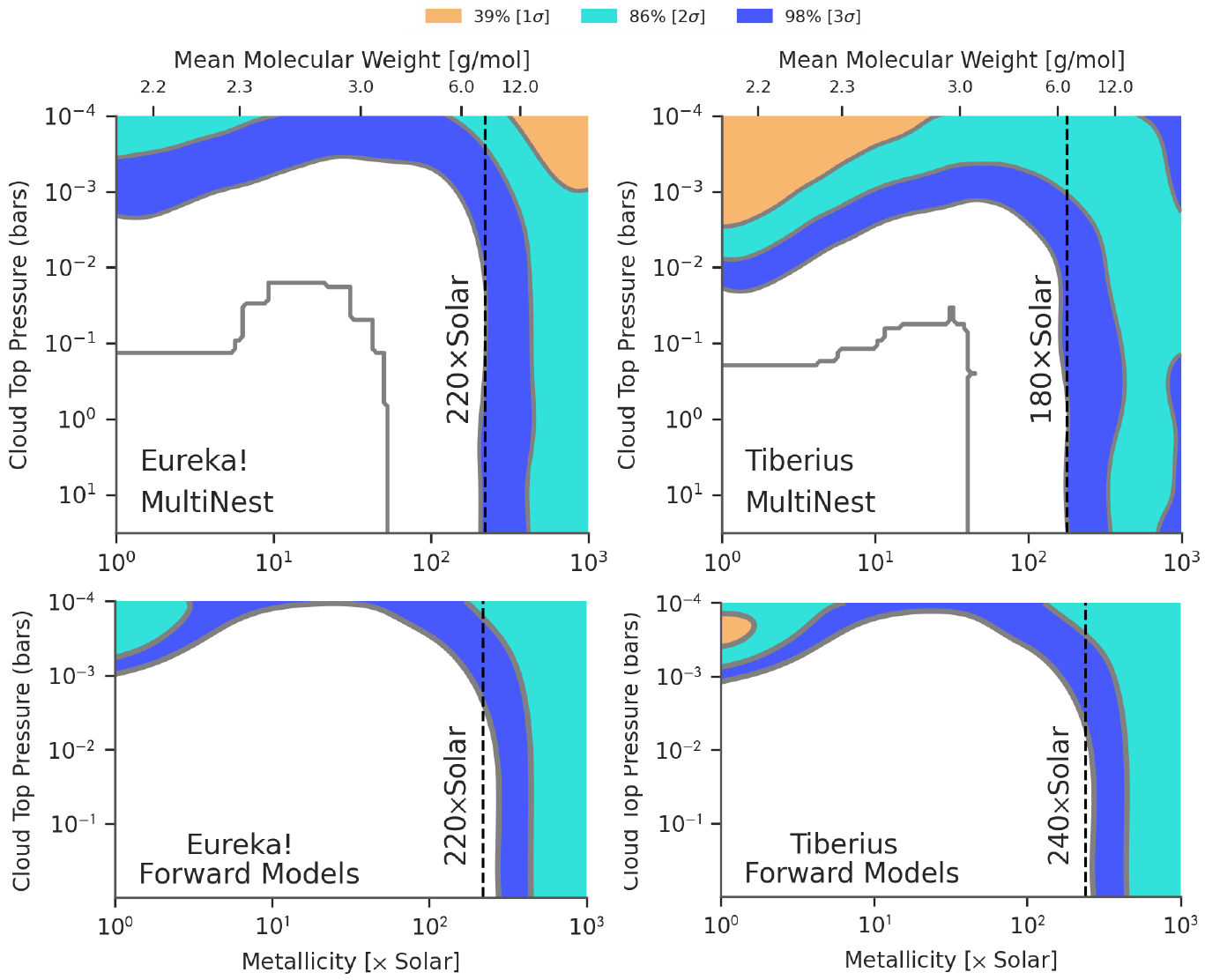}
    \caption{Two different methods to rule out parameter space in metallicity and opaque pressure level. Top row is when a Bayesian technique is used to retrieve the relevant physical parameters. Bottom row is when forward models are pre-generated and the $\sigma$-values are computed from a $\chi^2/N$ value. Filled contours indicate 39\%, 86\%, and 98\% probability regions, corresponding to two-dimensional 1, 2, and 3$\sigma$ thresholds. In the MultiNest results, the grey contour indicates the limit of computed posterior with regard to the scarcity of sampled points. Dashed black lines are used for reference. Using forward models tends to be slightly more optimistic. However, across both statistical methods and both data reduction methods, the average metallicity that can be ruled out at 3$\sigma$ is roughly 220$\times$Solar for opaque pressure levels greater than 10$^{-3}$ bars. }
    \label{fig:physical_heat}
\end{figure*}

Figure \ref{fig:physical_heat} shows the results in opaque pressure level-metallicity space for these two statistical analyses for the transmission spectra produced by \texttt{Eureka!} and \texttt{Tiberius}'s joint and averaged spectra, respectively. For the forward modeling approach, we use a two-dimensional cubic interpolation to get smooth contours in $\sigma$-space. For the nested sampling approach, we use the kernel density estimator available via the \texttt{arviz} \citep{kumar2019arviz} package to visualize this two-dimensional probability space in atmospheric metallicity and opaque pressure level.  For the \texttt{Tiberius} data reduction, the nested sampling approach provides a more conservative estimate for what parameter space can be ruled out at 3$\sigma$ (180$\times$Solar versus 240$\times$Solar) for pressures larger than 10$^{-3}$~bar. For the \texttt{Eureka!} data reduction, the two methods are comparable, both resulting in approximately 220$\times$Solar for pressures larger than 10$^{-3}$~bar. For pressures less than 10$^{-3}$~bar we start to lose model sensitivity even for Solar metallicities. For reference, a metallicity of 220$\times$Solar corresponds to a mean molecular weight of 7.7~g/mol. The exact mean molecular weight that can be ruled out, however, depends strongly on the chosen model. 

As described above, we were not able to conclusively diagnose the offsets in the spectra between visits. Indeed, determining whether/when different visits of the same target can be combined is an active area of exploration in the field. It is possible that such visit-to-visit differences are a result of stellar photosphere changes, that is, a visit-dependent transit light source effect (TLS; \citealt{Rackham2018}). In this case, for completeness, we also performed the same comparison of physical models with individual visit spectra from both \texttt{Eureka!} and \texttt{Tiberius}. The results are detailed in the Appendix, but in summary our conclusions do not change versus our comparison with the joint and weighted average spectra. 

\begin{figure*}
    \centering
    \includegraphics[width=0.9\textwidth]{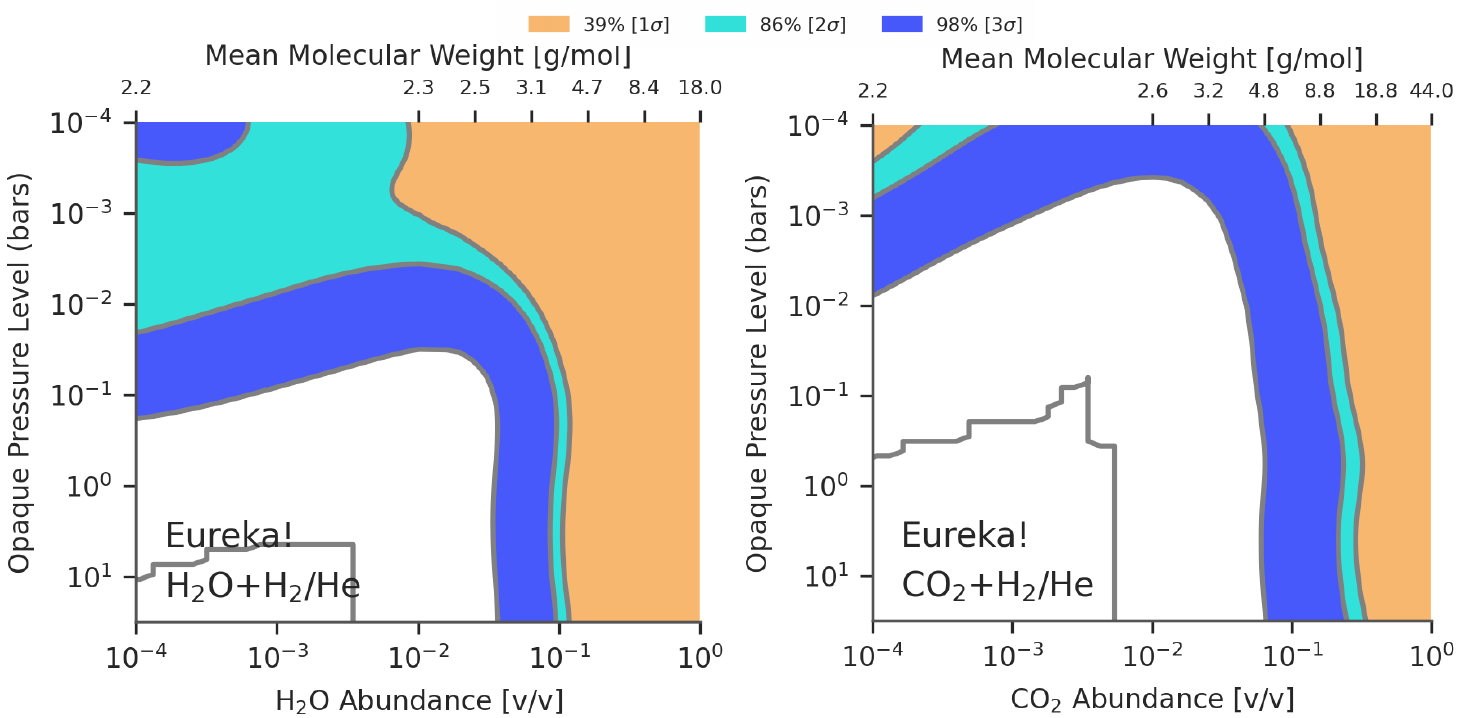}
    \caption{Similar as Figure \ref{fig:physical_heat} MultiNest results but ruling out parameter space in molecular abundance and opaque pressure level. Filled contours indicate 39\%, 86\%, and 98\% probability regions, corresponding to two-dimensional 1, 2, and 3$\sigma$ thresholds. On the right is the simple case of H$_2$O with an H$_2$/He background, and on the left is the same for CO$_2$ with an H$_2$/He background.
    Steam atmospheres are challenging to rule out, especially if clouds are present at pressures $<10^{-1}$~bar because of the lack of H$_2$O features in the NIRSpec/G395H bandpass. We can rule out CO$_2$ abundances of less than $\sim$10\% for a wider range of pressures levels ($<10^{-3}$~bar).}
    \label{fig:heat_molec}
\end{figure*}

We conduct two additional tests to understand the effect that the model has on the mean molecular weight that can be ruled out at 3$\sigma$. Using the nested sampling approach, we swap out the chemical equilibrium assumption for simple atmosphere mixtures: 1) H$_2$O with H$_2$/He and 2) CO$_2$ with  H$_2$/He. Instead of fitting for metallicity, we fit for the fractional abundance of either H$_2$O or CO$_2$ assuming a background of H$_2$/He. We keep the H$_2$/He fraction fixed at the Solar value \citep{Lodders2009}. Figure \ref{fig:heat_molec} shows the results of these tests for the \texttt{Eureka!}  jointly fit data. Steam atmospheres are more challenging to rule out when compared to the CO$_2$ models, because there are no complete (trough-to-peak-to-trough) H$_2$O spectral features in the NIRSpec/G395H bandpass. We can place 3$\sigma$ limits on the H$_2$O abundance in a pure H$_2$/He atmosphere of $>4$\% when the opaque pressure level is $>0.4$~bars. This corresponds to mean molecular weights of only 2.8~g/mol, a more pessimistic result than when compared to the case when a full chemical-equilibrium model is used (specified by the atmospheric metallicity). We also note that our result of 2.8 g/mol could be made more pessimistic by introducing featureless opacity source that has a lighter mean molecular weight than our H$_2$/He mixture. We can place 3$\sigma$ limits on the CO$_2$ abundance in a pure  H$_2$/He atmosphere of $>8$\% when the opaque pressure level is $>0.4$~mbars, which corresponds to a mean molecular weight of 5.5~g/mol. The CO$_2$ case is less susceptible to muting by clouds because its longer wavelength spectral feature at 4.3~$\mu$m becomes optically thick at pressures lower than the specified cloud deck. Though we do not show results for \texttt{Tiberius}, they are similar -- slightly more pessimistic for the H$_2$O mixture model ($>$2\% instead of $>$4\%) and slightly more optimistic for the CO$_2$ mixture model ($>$10\% instead of $>$8\%). Overall, the transmission spectra resulting from both reductions yield consistent model results. 

Ultimately, inferences on mean molecular weight are model dependent because they depend on what molecules are included in the model set, as well as the prominence of those molecules' absorption features in the observed bandpass. For the NIRSpec G395H transmission spectrum of TOI-776 c, chemical equilibrium models are the most optimistic, placing limits on atmospheric mean molecular weight of up  to 7.7~g/mol for opaque pressure levels larger than 10$^{-3}$~bar. In contrast, steam atmosphere models present the most observationally challenging case and result in limits of only 2.8~g/mol. This is expected since H$_2$O does not have strong molecular features in the NIRSpec/G395H bandpass, and therefore the relative feature strength as a function of mean molecular weight cannot be mapped. 

\begin{figure*}
    \centering
    \includegraphics[width=0.9\textwidth]{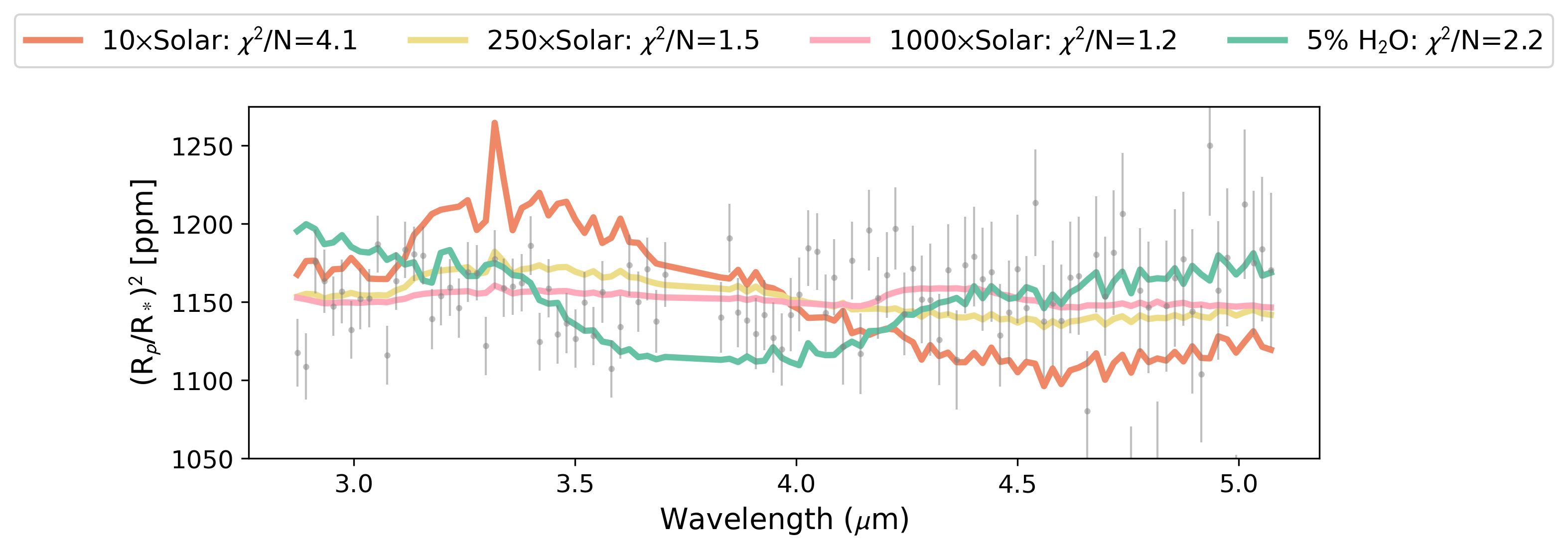}
    \caption{Transmission spectrum for the nominal \texttt{Eureka!} case (as detailed in Section~\ref{sec:eureka}) along with four example models depicting 10-1000$\times$, Solar metallicity (colored curves) and a steam atmosphere. The steam atmosphere is a 5\%H$_2$O, 95\% H$_2$/He mixture. The $\chi^2/N$ of each model is listed in the upper left. The computed $\chi^2/N$ do not include the NRS1/NRS2 detector offsets and error inflation terms that are used in the full retrieval.  Ultimately, the data permit  metallicities $>$220$\times$Solar and H$_2$O-rich atmospheres with more than 5\%H$_2$O. }
    \label{fig:spec}
\end{figure*}

In order to contextualize these results, we show representative spectra with the data in Figure \ref{fig:spec}. We include 10$\times$, 250$\times$ and 1000$\times$Solar metallicity spectra to demonstrate our loss of sensitivity around 220$\times$Solar. Note that in the chemical equilibrium models, CH$_4$ -- not CO$_2$ -- is expected to be the dominant carbon-bearing species for metallicities less than $\sim$250$\times$Solar. Only toward the 1000$\times$Solar case does the CO$_2$ start to show a spectral feature.  We also include a 5\%H$_2$O+95\% H$_2$/He mixture model to demonstrate our loss of sensitivity around 4\% H$_2$O. We do not include best-fit offsets or error inflation terms in the calculation of the displayed $\chi^2/N$, which could change the $\chi^2/N$ shown in the Figure. In principle, these terms would work to decrease the overall $\chi^2/N$  for a given case by increasing the noise term, and adjusting NRS1 and NRS2 to more closely match the model. However, the 1$\sigma$ confidence interval for the error inflation in the case of both model setups and data reduction methods is near zero (e.g., 0.2-7~ppm for \texttt{Tiberius}; 0.3-10~ppm for \texttt{Eureka!} in the metallicity retrieval case). Additionally, the 1$\sigma$ confidence interval for the offsets in the case of both model setups and data reduction methods is also near zero (2-14~ppm for \texttt{Tiberius}; 4-15~ppm for \texttt{Eureka!} in the metallicity retrieval case). Therefore, the spectra and $\chi^2/N$ shown in Figure \ref{fig:spec} are generally representative of the retrieval results.

\begin{figure*}
    \centering
    \includegraphics[width=0.9\textwidth]{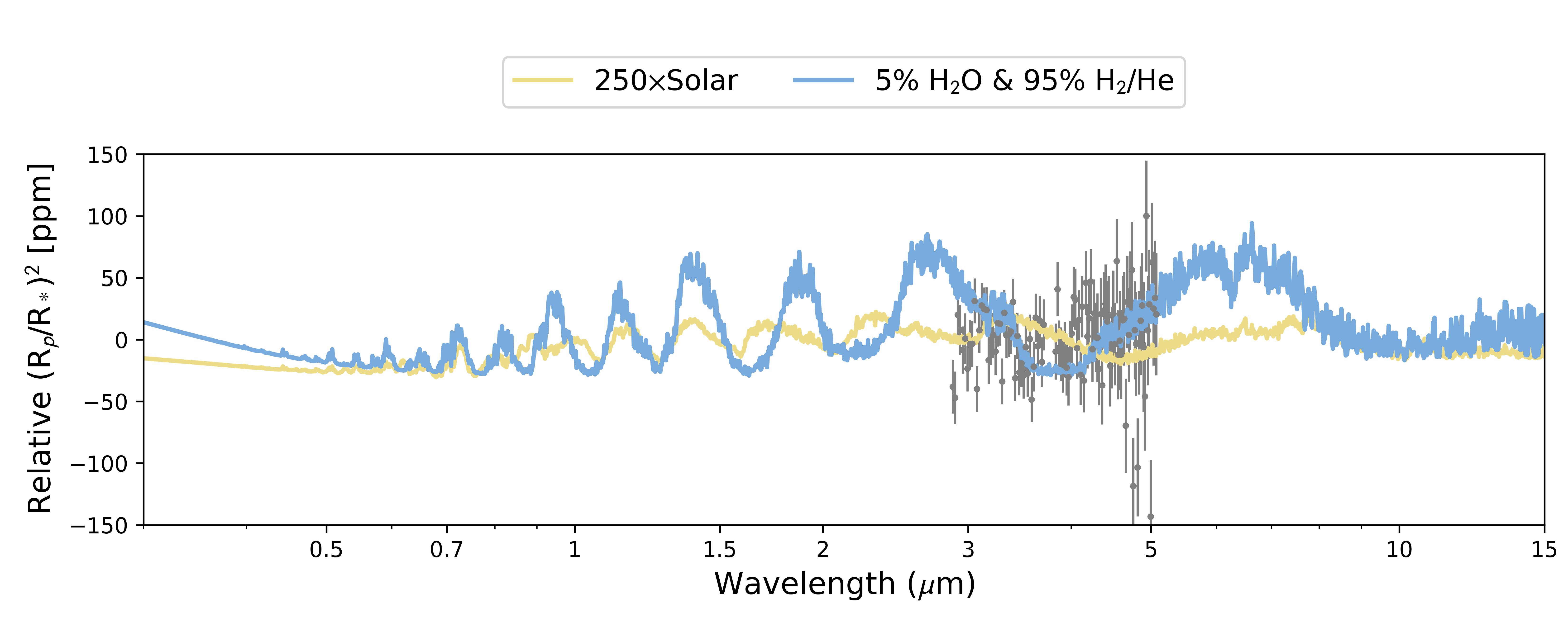}
    \caption{Transmission spectra for two edge case models, which shows how expanded wavelength coverage in transmission might aid in characterization of TOI-776~c's atmosphere. The 250$\times$Solar metallicity model (yellow) and the 5\% H$_2$O model (blue) both represent cases that are ruled out just below our 3$\sigma$ threshold. The nominal \texttt{Eureka!} joint reduction (as detailed in Section~\ref{sec:eureka}) is shown in grey. All data and models are shown relative to their mean transit depth between 3--5~$\mu$m. The 250$\times$Solar model does not contain larger features at shorter or longer wavelengths that would benefit from expanded wavelength coverage observations. However, if the spectrum is dominated by H$_2$O absorption alone, added wavelength coverage would help to characterize TOI-776~c's atmosphere.}
    \label{fig:specfuture}
\end{figure*}

%%%%%%%%%%%%%%%%%%%%%%%%%%%%%%%%%%%%%%%%%%%%%%%%%%%%%%%%%%%
\section{Discussion} \label{sec:discussion}
%%%%%%%%%%%%%%%%%%%%%%%%%%%%%%%%%%%%%%%%%%%%%%%%%%%%%%%%%%%

\subsection{TOI-776 c Atmosphere in Context with Other Similar Small Planets}

We do not detect any strong spectral features in the 3-5~$\mu$m transmission spectrum of the $\sim$2~R$_{\oplus}$, $T_{\rm{eq}}\sim 420$~K (assuming 0 Bond albedo) planet TOI-776 c. Our constraints on the mean molecular weight of the atmosphere depend on whether we assume chemical equilibrium or simple two-component mixtures. Our results suggest TOI-776 c could have a fairly high ($\gtrsim$200$\times$solar or 7.7 g/mol) mean molecular weight atmosphere, but could be as low as 2.8 g/mol depending on the adopted atmospheric composition.

Our findings contrast what was found for two other small planets, TOI-270~d and GJ 9827~d with molecular features detected in their JWST spectra. Given the greater similarity between TOI-776~c and TOI-270~d in terms of their host stars (M1V vs. M3V), temperatures ($T_{\rm{eq}}\sim$420 vs. $\sim$390~K), radii and masses (see Figures \ref{fig:mr_both} and \ref{fig:mr_teq}), here we focus on a comparison of these two planets. Both TOI-270~d and TOI-776~c fall within the $<500$~K regime suggested by \cite{Brande2024} to be relatively free of aerosols, based on their comparison of condensation cloud and hydrocarbon haze models to a sample of exo-Neptune transmission spectra taken with HST/WFC3's G141 grism. Both also fall into the density regime suggested by \cite{Luque2021} to represent planets with 50/50 water/rock (``water world'') compositions, but their bulk densities can be matched by multiple models (e.g., in Figure~\ref{fig:mr_both} left, TOI-776~c is colored with a $\sim$12\% water mass fraction). Atmospheric observations present a complementary opportunity to test this water-world hypothesis.

\citet{2024arXiv240303325B} and \citet{Holmberg2024A&A...683L...2H} reported multiple molecular species detections in the 0.6-5~$\mu$m JWST transmission spectrum of TOI-270~d, 
including CH$_4$, CO$_2$, and H$_2$O. 
The best match to the spectrum is a planet with 
a high mean molecular weight envelope ($\sim$5.5 g/mol, $\sim$225$\times$solar), which the authors infer is due to higher weight volatiles mixed with H$_2$/He in a miscible envelope. Combining this constraint with the mass and radius, the authors infer that only $\sim$10\% of the total mass of the planet is in the envelope, with the rest being in a rock/iron core. Thus, the results are not consistent with a 50/50 water/rock composition for this planet. 

The best fit atmospheric metallicity for TOI-270~d is $224\times$Solar with aerosol contribution only in the optical. The scale height that is associated with a $224\times$Solar metallicity atmosphere is allowed by the 3$\sigma$ limit of our detection threshold of 180-240$\times$Solar metallicity for TOI-776~c (depending on the data reduction and/or modeling method). Why does TOI-270~d show molecular features while TOI-776~c does not? 

\begin{figure}
   \centering
    \includegraphics[width=0.45\textwidth]{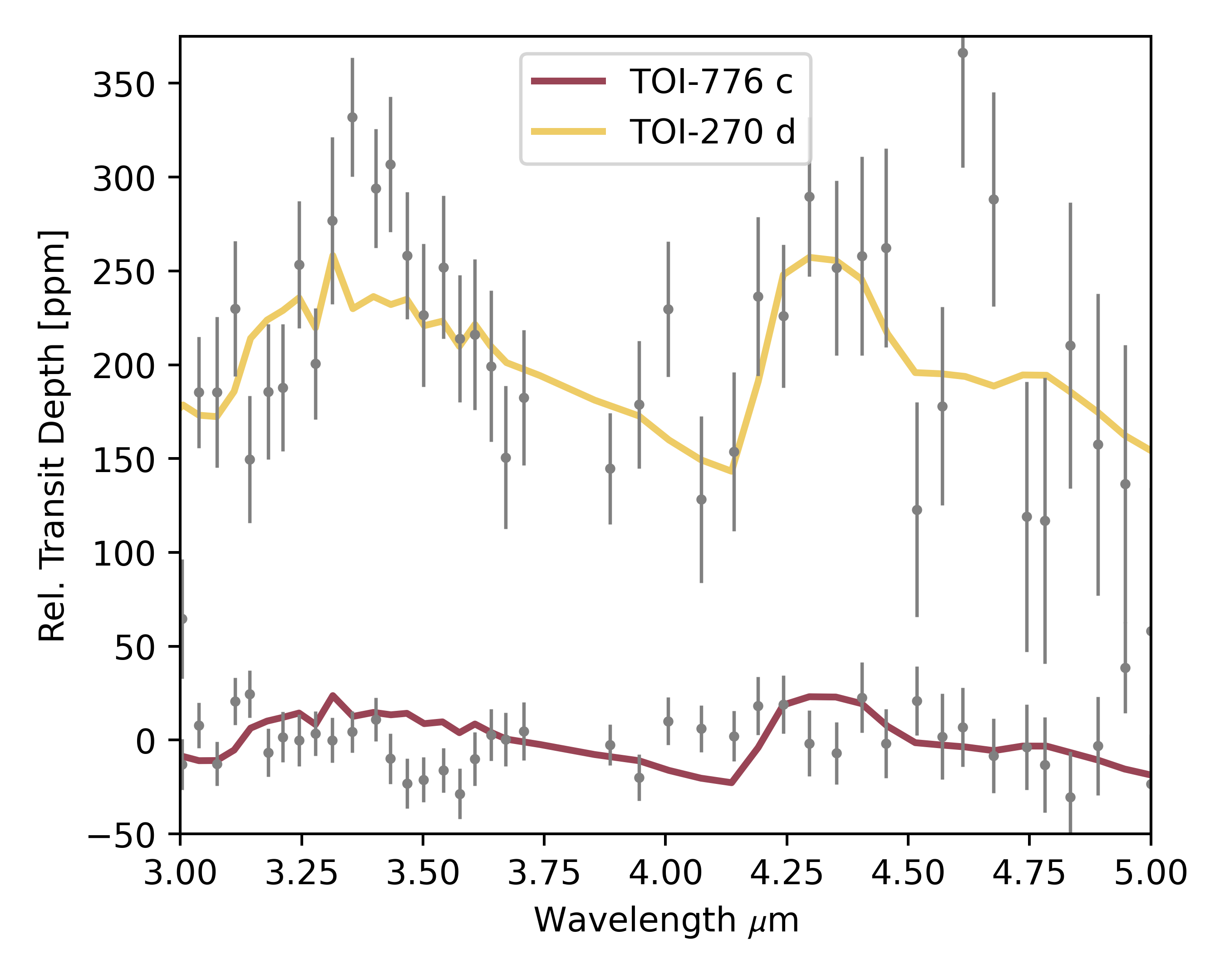}    
    \caption{Illustrative comparison of TOI-270 d and TOI-776 c. We use the retrieved composition from Table 2 in \citealt{2024arXiv240303325B} to create a spectrum of TOI-270 d. Using this same composition, we create a spectrum of TOI-776 c.} 
    \label{fig:t270d}
\end{figure}

If we consider the differences in the planetary and stellar properties, which contribute to the predicted feature size via the planet's scale height and planet-to-star ratio (i.e. surface gravity, planet and stellar radii, and equilibrium temperature), TOI-270~d is predicted to have a spectral feature size $\sim$1.7$\times$ that of TOI-776~c. To illustrate this, we obtain the median retrieved composition for TOI-270~d from Table 2 in \citealt{2024arXiv240303325B} and create spectra of both TOI-270~d and TOI-776 c using their independent planet and stellar properties, shown in Figure \ref{fig:t270d}. The peak-to-trough CO$_2$ amplitude is 80~ppm or TOI-270 d, compared to 46~ppm for TOI-776~c.   This comparison illustrates how even small differences in physical characteristics can impact our ability to characterize 1-3~R$_{\oplus}$ planets.

Given the remaining uncertainty regarding the nature of TOI-776~c, below we describe potential further investigations.   

\subsection{Future Prospects for Observing the TOI-776 System}

All of the targets in the COMPASS program were originally chosen due to their feasibility for radial velocity follow-up, so they are all bright and easily observable from the ground. This aspect of our program design makes ground-based atmospheric transmission spectroscopy an interesting potential avenue for further characterization. A fairly recent ground-based diagnostic of exoplanet atmospheres is the near-infrared triplet of He at 10830~\AA\,.  There is still some uncertainty around trends in He 10830~\AA\ detections -- see the discussion in \cite{Alam2024} -- but it is a potential diagnostic of mass-loss and metallicity (between 10-200$\times$solar) in small planets and complementary to the NIR observations with JWST. 
Most previous He {\sc i} 
detections have been in young and/or larger planets \citep[e.g.,][]{Spake2018,Allart2018,Kirk2020,Zhang2023a}, 
but there have been two recent and somewhat surprising detections in smaller, ``mature'' planets \citep{Zhang2023b,Zhang2024}. These detections include TOI-836~c, another COMPASS target published in \cite{Wallack2024} that showed no strong molecular absorption features in the 3-5~$\mu$m spectrum. Our optimistic atmospheric metallicity constraint for TOI-776~c suggests that we should not detect excess He absorption, which is in line with theoretical predictions that planets with higher metallicity atmospheres are able to withstand photoevaporation at short orbital periods \citep{Owen2018}. However, our data still allow for scenarios wherein TOI-776~c has a lower metallicity atmosphere and high altitude clouds/hazes (see Figures~\ref{fig:physical_heat} and \ref{fig:heat_molec}). Thus a robust He non-detection 
could place informative upper limits on the atmospheric mass loss rate of TOI-776~c and provide observational tests on the effect of planetary atmospheric metallicity on escape processes \citep{Owen2018, Piaulet-Ghorayeb2024}. 

However, caution is warranted with interpreting He 10830~\AA\, observations because the line can be depopulated by a variety of mechanisms, such as stellar winds and magnetic field effects \citep[e.g.,]{Carolan2020,Schreyer2024,Alam2024}. Thus, escape may be occurring, but helium would not be in the metastable state and therefore unobservable at 10830~\AA. An alternative window into the upper atmosphere may therefore be needed, such as the Ly-$\alpha$ line that has been a productive method for detecting atmospheric escape with HST, albeit also in younger and/or larger planets \citet{Kulow2014,Ehrenreich2015,Lavie2017,Bourrier2018,Zhang2022}. For TOI-776, the large systemic velocity of the star ($\sim$49~km\,s$^{-1}$) actually makes this system a favorable target for Ly-$\alpha$ observations. Although we point out that  in high metallicity atmospheres, the ion fraction of H in the outflow is expected to be much higher \citep{Piaulet-Ghorayeb2024}, which could make Ly-$\alpha$ detections difficult. 

Future observations with JWST could provide further insight into the nature of TOI-776~c's atmosphere through eclipse observations or additional transmission visits. We can explore these options by leveraging a model at the limit of our data (250$\times$Solar metallicity). This scenario is also interesting to explore given its similarity to TOI-270~d. In thermal emission, if TOI-776~c had no atmosphere, we would expect to detect a signal of 90~ppm at 15$\mu$m (observable by MIRI photometry) assuming the parametrization for no heat redistribution derived in \citet{Koll2022ApJ...924..134K}. Given the bulk density constraint on the planet and its temperature (see Figure \ref{fig:mr_teq}), we do not expect a bare rock planet. 
If instead TOI-776~c were to have a thick 250$\times$Solar metallicity atmosphere, assuming full heat redistribution, we would expect a smaller signal of $\sim50$~ppm. Assuming a Phoenix stellar grid model \citep{phoenix1995ApJ...445..433A} and an integration time of 2$\times$transit duration, the JWST ETC \citep{2016SPIE.9910E..16P} predicts a precision of 31~ppm with one eclipse of MIRI F1500W. Therefore, roughly 4 eclipses would be needed to detect a $\sim50$~ppm signal. Although this seems observationally attainable, more in-depth theoretical modeling is needed in order to determine exactly what information content could be derived from MIRI photometry alone. 
In transmission, the question remains as to whether or not added wavelength coverage would be preferable to added visits in NIRSpec/G395H. We can partially address this by  looking at the expanded wavelength coverage for the 3$\sigma$ edge case scenarios, shown in Figure \ref{fig:specfuture}. For the 250$\times$Solar metallicity model, the largest feature is CH$_4$ in the wavelength ranges studied in this analysis.   
Two additional visits with NIRSpec/G395H would enable us to rule out scenarios up to 300$\times$Solar metallicity, which would allow for TOI-270 d's atmospheric metallicity to be confidentially ruled out. Adding shorter or longer wavelengths  with, for example, NIRISS/SOSS or MIRI LRS would be more favorable to rule out parameter space in mean molecular weight if TOI-776 c had an atmosphere of pure H$_2$O and H$_2$/He.

%%%%%%%%%%%%%%%%%%%%%%%%%%%%%%%%%%%%%%%%%%%%%%%%%%%%%%%%%%%
\section{Summary \& Conclusions} \label{sec:summary}
%%%%%%%%%%%%%%%%%%%%%%%%%%%%%%%%%%%%%%%%%%%%%%%%%%%%%%%%%%%

In this paper, we present the JWST NIRSpec/G395H transmission spectrum of the $\sim$420~K, $\sim$2~R$_{\oplus}$, $\sim$7~M$_{\oplus}$ planet TOI-776~c, gathered from two visits. 
This planet 
is among a limited but very interesting group of planets with bulk densities consistent with a $\sim$50\% water composition \citep{Luque2021}, perhaps constituting a different population besides gas dwarfs and airless rocks (although we note in the case of TOI-776~c, the water would likely be in super-critical/vapor form, decreasing the inferred water mass fraction as shown in Figure \ref{fig:mr_both}).  
The planet also falls within the radius valley that divides the small ($R_p < 3.5 R_{\oplus}$) planet population \citep{Fulton2018}, a part of parameter space suggested to represent planets transitioning from having envelopes still possessing primordial H$_2$/He to those stripped of primordial gasses \citep[e.g.,][]{Owen2017,Rogers2023}. Determining the nature of the atmosphere of TOI-776~c is thus relevant to theories of small planet formation and evolution.

Through two independent reductions of the two transits we find fairly consistent results (see Figure \ref{fig:visit_spec}). Our analysis of the spectra with non-physical models shows some slight differences in model preferences (e.g., zero-sloped line versus offset between detectors), but the combined-visit spectra show no indication of significant features, including at 3.3~$\mu$m around CH$_{4}$'s prominent spectral feature. Our analysis of the spectra with physical models (forward models as well as Bayesian-based retrievals) allows us to rule out atmospheres with metallicities $\lesssim$180-240$\times$solar to $3\sigma$, depending on the reduction pipeline and modeling approach. Given the small visit-to-visit differences, we also investigated the individual visit spectra and find similar constraints (see the Appendix). We find that atmospheric mean molecular weight inferences are model dependent, such that chemical equilibrium models give a more optimistic constraint (7.7 g/mol for pressures $>10^{-3}$ bar). Two-component chemical mixture models of H$_2$O or CO$_2$ with H$_2$/He as the background give more pessimistic limits of 2.8 g/mol and 5.5 g/mol, respectively. 

We find no statistically significant features in the transmission spectrum of TOI-776~c, which contrasts the fairly similar planet TOI-270~d that does show molecular features in transmission. We find that the atmospheric composition derived for TOI-270~d  \citep{2024arXiv240303325B} is just within the 3$\sigma$ limit of our detection threshold. Thus with these TOI-776~c observations we may be brushing right up against the combination of planet and system parameters that result in detectable features in small planet spectra. We suggest a few  
avenues for further characterizing this planet's atmosphere, but acknowledge that robustly distinguishing high metallicity from cloudy/hazy atmospheres will likely remain a challenge even in the JWST era. A large sample of small planet atmospheres will be needed to understand the implications of
even small system parameter differences, and whether we are seeing evidence of multiple populations.  

\appendix 

\section{Atmospheric Constraints from Physical Models of Individual Visit Spectra}

\begin{figure*}
    \centering
    \includegraphics[width=0.9\textwidth]{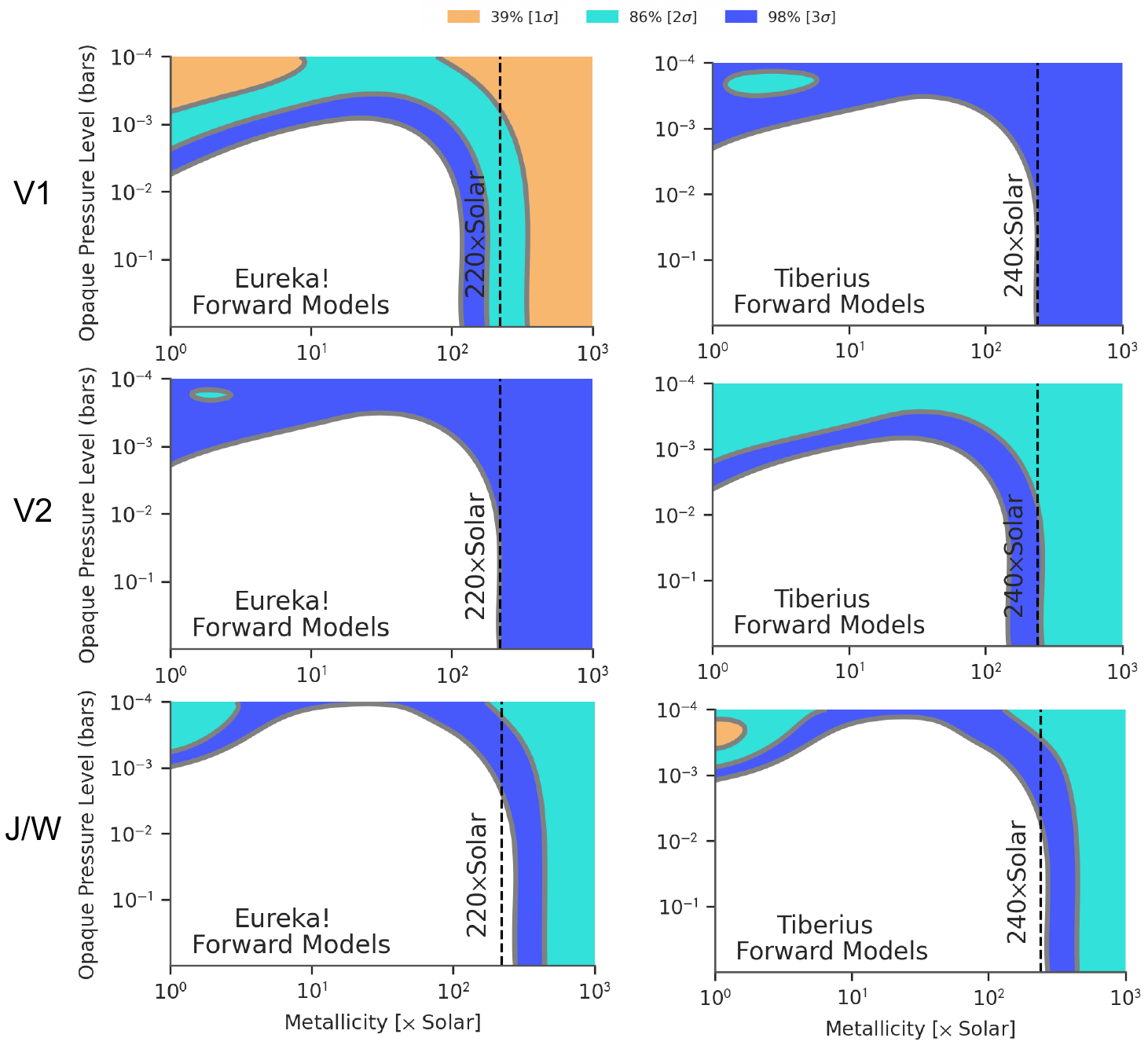}
    \caption{Here we show a different version of Figure~\ref{fig:physical_heat}, wherein the method to rule out parameter space in metallicity and opaque pressure level is held constant (same as bottom row of Figure~\ref{fig:physical_heat}) but we change the visits used. The top row shows just visit 1, the middle row shows just visit 2, and the bottom row shows (again, for ease of comparison) the joint and weighted average spectra for \texttt{Eureka!} and \texttt{Tiberius}, respectively. Filled contours indicate 39\%, 86\%, and 98\% probability regions, corresponding to two-dimensional 1, 2, and 3$\sigma$ thresholds.}
    \label{fig:physical_heat_visits}
\end{figure*}

Given the offsets between visits, we wanted to test whether our metallicity constraints were significantly different between visits or comparing visits with the joint (\texttt{Eureka!}) or weighted average (\texttt{Tiberius}) spectrum. As shown in Figure~\ref{fig:physical_heat_visits}, the weighted average for \texttt{Tiberius} and the joint fit for \texttt{Eureka!} are not enabling us to rule out significantly more of the parameter space -- the constraints for \texttt{Tiberius} visit one are similar to the weighted average, and the constraints for \texttt{Eureka!} visit 2 are similar to the joint fit. Thus our conclusions are not affected.

%%%%%%%%%%%%%%%%%%%%%%%%%%%%%%%%%%%%%%%%%%%%%%%%%%%%%%%%%%%

\begin{acknowledgments}
\noindent The data products for this manuscript can be found at the following Zenodo repository: \href{https://doi.org/10.5281/zenodo.14841601}{https://doi.org/10.5281/zenodo.14841601}.

This work is based on observations made with the NASA/ESA/CSA James Webb Space Telescope. The data were obtained from the Mikulski Archive for Space Telescopes at the Space Telescope Science Institute, which is operated by the Association of Universities for Research in Astronomy, Inc., under NASA contract NAS 5-03127 for JWST. These observations are associated with program \#2512. Support for program \#2512 was provided by NASA through a grant from the Space Telescope Science Institute, which is operated by the Association of Universities for Research in Astronomy, Inc., under NASA contract NAS5-03127.

This work is funded in part by the Alfred P. Sloan Foundation under grant G202114194.

Support for this work was provided by NASA through grant 80NSSC19K0290 to JT and NLW. %%MAKE SURE NLW GETS IN PROOFS%%

This work benefited from the 2022 and 2023 Exoplanet Summer Program in the Other Worlds Laboratory (OWL) at the University of California, Santa Cruz, a program funded by the Heising-Simons Foundation. 

JK acknowledges financial support from Imperial College London through an Imperial College Research Fellowship grant.

AA is supported by NASA’S Interdisciplinary Consortia for Astrobiology Research (NNH19ZDA001N-ICAR) under grant number 80NSSC21K0597.

S.E.M. is supported by NASA through the NASA Hubble Fellowship grant HST-HF2-51563 awarded by the Space Telescope Science Institute, which is operated by the Association of Universities for Research in Astronomy, Inc., for NASA, under contract NAS5-26555.

%%remember our collaboration agreement and make sure everyone's contributions are acknowledged 
%https://compass-jwst.github.io/collab.html
Co-Author contributions are as follows: JT wrote significant parts of the paper text, produced the introduction and observation section figures and tables, and contributed leadership as Co-PI. NEB performed the atmospheric modeling, producing both the non-physical and physical constraints for TOI-776~c as well as the comparison to TOI-270~d, described and tabulated these results in the paper, wrote significant parts of the paper text, and contributed leadership as PI. NLW and JK provided independent data reduction, data products, spectra, and associated text. NFW provided atmosphere modeling code and computed JWST ETC calculations. TAG performed independent fits to the white light curves including a GP component, which help contextualize the original fits. MKA provided visualization code and contributed text to the He {\sc i} discussion. AA provided interior modeling code and visualization code. AW provided statistics expertise to interpretation of spectra. All co-authors provided feedback on the paper draft and/or during team meetings.  

We thank the referee for their thorough comments that improved the paper.
\end{acknowledgments}

\software{mc3 \citep{Cubillos2017}, picaso \citep{2019ApJ...878...70B}, photochem \citep{Wogan2023}, numpy \citep{walt2011numpy}, scipy \citep{2020SciPy-NMeth}, ultranest \citep{ultranest2021JOSS....6.3001B}, arviz \citep{kumar2019arviz}, emcee \citep{Foreman-Mackey2013}}

\bibliography{references}{}
\bibliographystyle{aasjournal}

\end{document}